\documentclass[%
 reprint,
 amsmath,amssymb,
 aps,
pra,
floatfix
]{revtex4-2}

\usepackage{xcolor}

\usepackage{graphicx}
\usepackage{dcolumn}
\usepackage{bm}
\usepackage[mathlines]{lineno}

\begin{document}


\title{Selection bias effects on high-$p_\mathrm{T}$ yield and correlation measurements in Oxygen+Oxygen collisions}%

\author{JaeBeom Park}
 \email{jaebeom.park@colorado.edu}
\author{J.L. Nagle}%
 \email{jamie.nagle@colorado.edu}
\author{Dennis V. Perepelitsa}%
 \email{dvp@colorado.edu}
\affiliation{Department of Physics, University of Colorado Boulder, Boulder, CO, USA}

\author{Sanghoon Lim}
\email{shlim@bnl.gov}
\affiliation{Department of Physics, Pusan National University, Busan, South Korea}

\author{Constantin Loizides}
\email{constantin.loizides@cern.ch}
\affiliation{Department of Physics, Rice University, Houston, TX, USA}
\affiliation{CERN, Geneva, Switzerland}


\date{\today}

\begin{abstract}
Oxygen+Oxygen (O+O) collisions at RHIC and the LHC offer a unique experimental opportunity to observe the onset of jet quenching in intermediate relativistic collision systems. As with the smaller proton--nucleus or larger nucleus--nucleus systems, measurements of centrality-selected high-$p_\mathrm{T}$ processes in O+O collisions are expected to be sensitive to selection bias effects, which will be necessary to quantify or mitigate before a definitive conclusion on the presence of jet quenching. Using two Monte Carlo heavy-ion event generators, we provide a survey of centrality bias effects on high-$p_\mathrm{T}$ yield and correlation measurements.
Some highlights of our findings include that (1) bias factors for the accessible kinematic range at RHIC show a non-trivial $p_\mathrm{T}$ dependence, compared to a negligible one at the LHC given the smaller accessible Bjorken-$x$ range, (2) centrality definitions based on multiplicity are less sensitive to bias effects than those based on the transverse energy, (3) the \textsc{Angantyr} generator gives qualitatively similar but larger-magnitude bias factors than \textsc{Hijing}, and (4) correlation measurements have a much smaller sensitivity to bias effects than do yield measurements.
The findings here are intended to guide the experimental design and interpretation of O+O jet quenching and other hard-process measurements.
\end{abstract}


\maketitle


\section{Introduction}

Oxygen+Oxygen (O+O) collisions at the Relativistic Heavy Ion Collider (RHIC) and the Large Hadron Collider (LHC) have the potential to uniquely address a number of open scientific questions in heavy-ion physics~\cite{Brewer:2021kiv}. One key question concerns the possible onset of significant parton-medium interactions, i.e., jet quenching, in proton-nucleus ($p$+A) collisions, that are readily observed in large nucleus--nucleus (A+A) collision systems which produce Quark-Gluon Plasma (QGP)~\cite{Cunqueiro:2021wls}. Despite broad, robust evidence that $p$+A collisions exhibit many signatures of QGP formation~\cite{Loizides:2016tew,Nagle:2018nvi}, a definitive observation of jet quenching in these systems remains elusive, with many measurements setting strict constraints on its possible magnitude~\cite{ALICE:2017svf,ATLAS:2022iyq,CMS:2025jbv}. Searches for jet quenching in central $p$+A collisions are further exacerbated by the extreme asymmetry of the collision system and the lone proton, which introduces experimental and theoretical modeling challenges~\cite{ATLAS:2023zfx,STAR:2024nwm,Perepelitsa:2024eik}. By contrast, O+O collisions allow for a unique experimental opportunity to explore the onset of jet quenching effects in symmetric, intermediate-sized systems~\cite{Katz:2019qwv,Huss:2020dwe,Huss:2020whe,Zakharov:2021uza,Ke:2022gkq,Ogrodnik:2024qug,vanderSchee:2025hoe} which span a similar multiplicity range as $p$+A collisions but are more experimentally and theoretically tractable. 

RHIC provided O+O collisions at $\sqrt{s_\mathrm{NN}} = 200$~GeV to the STAR experiment~\cite{STAR:2025ivi} in May 2021 and sPHENIX~\cite{Belmont:2023fau} has similarly proposed a short O+O run at the conclusion of RHIC Run-25. In July 2025, the LHC plans to provide O+O collisions at $\sqrt{s_\mathrm{NN}} = 5.36$~TeV to the experiments at that facility. Interestingly, O+O will serve as the first collision system besides proton--proton ($p$+$p$) for which data with the same nuclear species will be available at both RHIC and the LHC.

In large collision systems, jet quenching is observed in a variety of ways, such as from the suppression of high-$p_\mathrm{T}$ jets and hadrons from hard-scattered partons traversing the QGP or from a distortion of kinematic correlations between high-$p_\mathrm{T}$ objects such as photon+jet ($\gamma$+jet) pairs. As preparation for these measurements in the O+O system, recent efforts have focused on quantifying the possible contribution to experimental measurements from other physics effects, such as the modification of nuclear parton densities and uncertainties in perturbative-QCD calculations~\cite{Gebhard:2024flv}, and on uncertainties inherent in the choice of $p$+$p$ reference for the O+O system at the LHC~\cite{Brewer:2021tyv}.

Jet quenching effects are expected to be strongest in the events with the largest nuclear overlap, and such events are traditionally selected via their centrality~\cite{ALICE:2014xsp}, an experimentally-defined measure of their overall event activity understood to be correlated with the underlying event geometry. 
In this paper, also as preparation for the interpretation of first O+O measurements, we quantify the impact of experimental selection biases on measurements of high-$p_\mathrm{T}$ yields and correlations in centrality-selected O+O collisions at RHIC and the LHC. If not accounted for such biases could, for example, result in an increase of the measured nuclear modification factor, thus decreasing the visible jet quenching signature and leading to an incorrect interpretation of the physics of these systems.
Since possible jet quenching effects in O+O collisions are expected to be modest and potentially of a similar magnitude to such bias effects, it will be critical to understand or mitigate them to allow for the unambiguous observation of jet quenching.

As an alternative strategy, measurements performed in minimum-bias O+O collisions will be free from such selection effects, but will come with other challenges, such as a weaker jet quenching signal than in the most central collisions, the potential contamination from electromagnetic interactions~\cite{Klein:2020fmr}, and Oxygen beam transmutation backgrounds~\cite{Bruce:2021hjk,Nijs:2025qxm}. As a separate strategy, geometric classes could also be selected based on the number of spectator neutrons measured in Zero Degree Calorimeters~\cite{ALICE:2014xsp,ALICE:2019fhe,ATLAS:2022iyq}, however this may be difficult in the given system due to the much smaller dynamic range of neutrons in O+O collisions compared to, e.g., Pb+Pb. For the most complete view of jet quenching effects, it will be necessary to pursue a variety of strategies, including the centrality-selected measurements discussed here. 

The quantitative effect of multiplicity selection biases on hard process measurements in $p$+A and similar (deuteron--nucleus, etc.) collisions have been studied at RHIC and the LHC~\cite{PHENIX:2013jxf,ALICE:2014xsp,Perepelitsa:2014yta,Loizides:2017sqq}, with the general feature that upward multiplicity fluctuations in events with a moderate-$p_\mathrm{T}$ hard process causes the yield of such processes to be enhanced in central, or high-multiplicity, selected events and to be suppressed in peripheral, or low-multiplicity, selected events. This may be understood as arising from the positive correlation between hard scattering rates and soft particle production in individual nucleon--nucleon collisions, potentially due to an increased rate of multi-parton interactions (MPIs)~\cite{Sjostrand:1986ep} or an impact parameter picture of nucleon--nucleon collisions~\cite{Jia:2009mq,Frankfurt:2010ea}. Such hard--soft correlations are also general features of $p$+$p$ collisions~\cite{CDF:2001onq}.

Notably, the multiplicity bias effect is also observed in peripheral A+A collision data~\cite{ALICE:2018ekf,CMS:2021kvd,ALICE:2024yvg}, strongly suggesting that it will be relevant in intermediate-sized systems such as O+O as well. Ref.~\cite{Loizides:2017sqq} provided an initial calculation of such effects within a Monte Carlo (MC) model of \textsc{Pythia}~6.4~\cite{Sjostrand:2006za} $p$+$p$ events overlaid on top of \textsc{Hijing}~\cite{Wang:1991hta} event collision geometries, and determined $p_\mathrm{T}$-independent bias factors for charged particles in O+O collisions at the LHC. 

In this paper, we quantify the bias factors for a variety of yield and correlation measurements accessible to the experiments in O+O collisions, over a wide kinematic range where the magnitudes and even signs of bias effects may vary, within the \textsc{Hijing} and recent \textsc{Angantyr}~\cite{Bierlich:2018xfw} MC event generators, and at both RHIC and LHC energies. We further separate the underlying origin of the bias into that arising due to multiplicity fluctuations and the direct bias from associated jet production, and we investigate the sensitivity of different centrality definitions. The results here are intended to provide quantitative guidance in the design of experimental measurements and in their interpretation, particularly regarding the possible observation of, or upper limit on, jet quenching effects.

\section{Method}

Two MC heavy-ion event generators were used to evaluate bias factors on high-$p_\mathrm{T}$ yield and correlation measurements in centrality-selected O+O collisions, with the comparison between them serving as a key test of the model dependence. \textsc{Hijing} v1.383 was run with final-state radiation turned off. \textsc{Angantyr} was run within the \textsc{pythia} 8.314 release~\cite{Bierlich:2022pfr} in minimum-bias mode with its default settings. For both generators, the default description of the Oxygen nucleus wavefunction for the Glauber nucleon modeling was used. For the bias effects studied here, different descriptions of the Oxygen nucleus~\cite{Lim:2018huo} are not expected to lead to qualitatively different conclusions.

Our study examines possible centrality definitions which are approximately matched to those used, or expected to be used, by the experiments. We examine centrality definitions based on the total charged-particle multiplicity ($N_\mathrm{ch}$) and, separately, on the sum of the transverse energy ($\Sigma E_\mathrm{T}$), for particles within some defined pseudorapidity range. Comparing the bias factors for $N_\mathrm{ch}$ vs. $\Sigma E_\mathrm{T}$ centrality definitions can give a handle on, for example, the susceptibility to a bias from other jets in the event associated with the hard scattering.

At RHIC, centrality is defined in the region $2.1 < \left|\eta\right| < 5.1$, corresponding to the acceptance of the STAR Event Plane Detector (EPD) scintillator~\cite{Adams:2019fpo} which principally measures $N_\mathrm{ch}$. At the LHC, the centrality-defining region is $3.2 < \left|\eta\right| < 4.9$, corresponding to the acceptance of the ATLAS Forward Calorimeter (FCal), where the $\Sigma E_\mathrm{T}$ has been used in previous ATLAS measurements in small and large systems~\cite{ATLAS:2015hkr}. The most commonly-used centrality definition in CMS is the $\Sigma E_\mathrm{T}$ in the hadron forward (HF) calorimeter which sits in a slightly more forward pseudorapidity range, $4 < \left|\eta\right| < 5$~\cite{CMS:2018xfv}. The ALICE experiment typically uses centrality measures which are in a similar forward pseudorapidity range but which are sensitive principally to $N_\mathrm{ch}$, for example the FT0A detector covering the range $3.5 < \eta < 4.9$ in Run-3~\cite{ALICE:2025cjn} or VZERO-A covering $2.8 < \eta < 5.1$ in Run-1/Run-2~\cite{ALICE:2013hur}. 

Similarly, the kinematic selections on the high-$p_\mathrm{T}$ final states explored here are guided by those used in preliminary measurements by STAR at RHIC or in selections commonly used in small systems by the LHC experiments. In particular, the pseudorapidity ranges of the centrality-defining region and the region in which high-$p_\mathrm{T}$ final states are measured are important to model realistically, since selection bias effects are known to be sensitive to the separation between these. For the STAR measurements at RHIC~\cite{Zhang:2025cgr}, we evaluate measurements of charged-hadron yields within $\left|\eta\right| < 0.5$, $R=0.2$ charged-particle jets within $\left|\eta\right| < 1.3$, and hadron-triggered semi-inclusive recoil jets where the trigger hadron has $p_\mathrm{T} > 7$~GeV and is within $\left|\eta\right| < 1.5$ and the associated jets are within $\left|\eta\right| < 1.3$ and are azimuthally correlated with the hadron via $\Delta\phi > 3\pi/4$. For measurements at the LHC, we evaluate charged-hadron yields within $\left|\eta\right| < 2.5$ and jet yields within $\left|\eta\right| < 2.1$, matching that in recent ATLAS measurements~\cite{ATLAS:2022kqu,ATLAS:2022vii}. We also evaluate measurements of hadron--jet correlations for trigger hadrons with $p_\mathrm{T} = 12$--$50$~GeV within $\left|\eta\right| < 0.9$ and associated charged-particle jets with $\left|\Delta\phi - \pi\right| < 0.6$ and in the region $\left|\eta\right| < 0.7$, identical to previous ALICE measurements in small systems~\cite{ALICE:2017svf}.

The following procedure was used to determine the bias factors, and is similar that used in Ref.~\cite{PHENIX:2013jxf}. Separately for each collision energy and centrality definition, 10\%-wide centrality bins were defined by sorting the given centrality quantity over all events. The distribution of the number of nucleon--nucleon collisions, $N_\mathrm{coll}$, was recorded for each such event class. The hard-process yield or the per-trigger correlated yield were determined in each centrality class. These yields represent what would be measured under an experimentally-defined centrality selection, including the bias selection effect. Then, for each centrality class, the yields were determined again but in the set of all events, weighted event-by-event to reproduce the centrality-selected $N_\mathrm{coll}$ distribution. The latter yields thus represent what would be produced from an ensemble of events with the same $N_\mathrm{coll}$, but agnostic as to the produced centrality signal and thus unbiased by such effects. The ratio of these yields thus quantifies the change in the yield induced by the multiplicity bias effect. Notably, the ratio is not sensitive to any physics in the given generator itself, which may modify the $N_\mathrm{coll}$-dependence of particle production (e.g., shadowing), and isolates primarily the multiplicity bias effect.

\begin{figure}[!h]
\includegraphics[width=\linewidth]{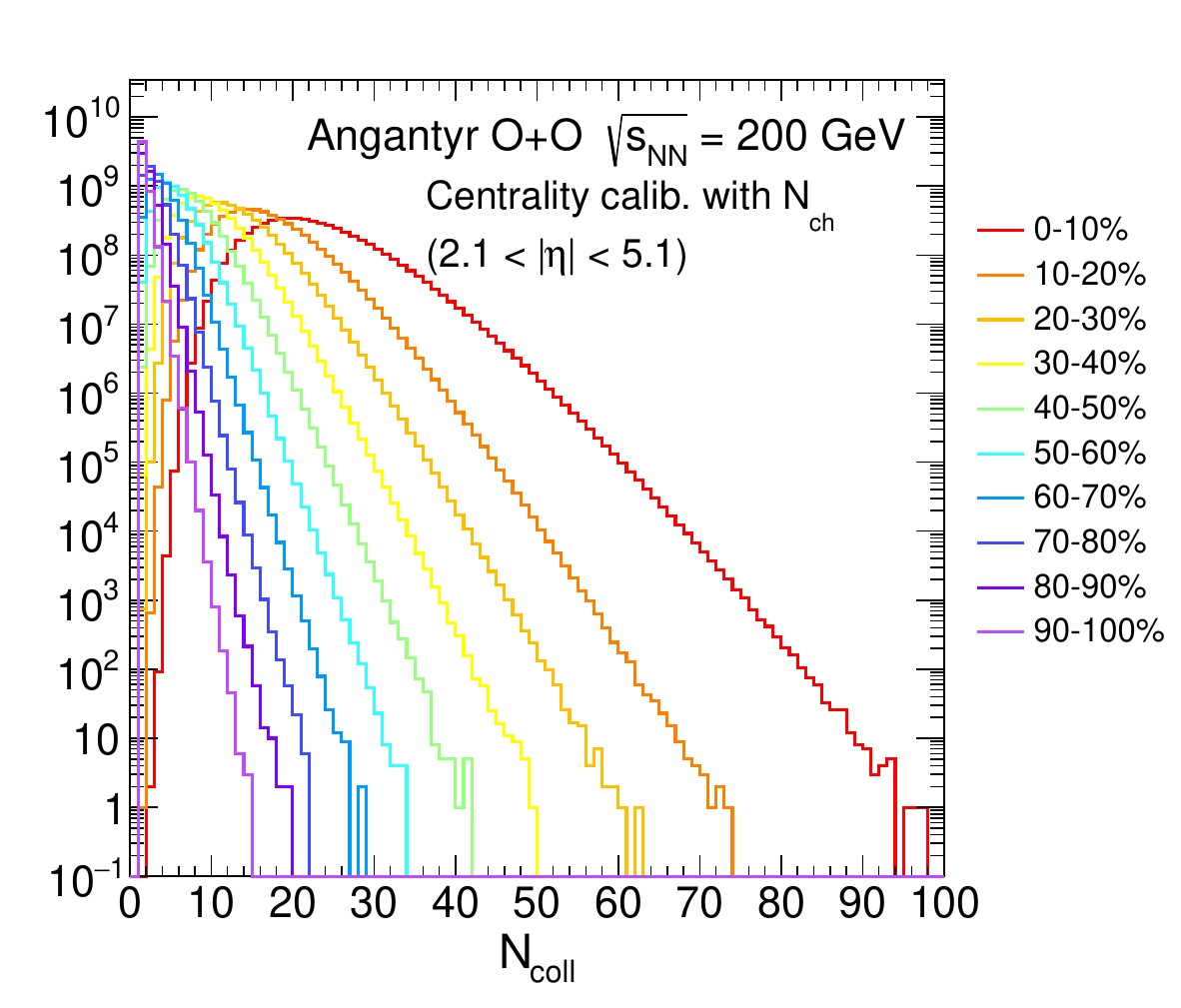}
\caption{\label{fig:ncoll} $N_\mathrm{coll}$ distributions in centrality-selected O+O events in the \textsc{Angantyr} heavy-ion event generator. Results are shown for RHIC with a forward charged-particle multiplicity-based definition, $N_\mathrm{ch}$.}
\end{figure}

\begin{figure}[!t]
\includegraphics[width=\linewidth]{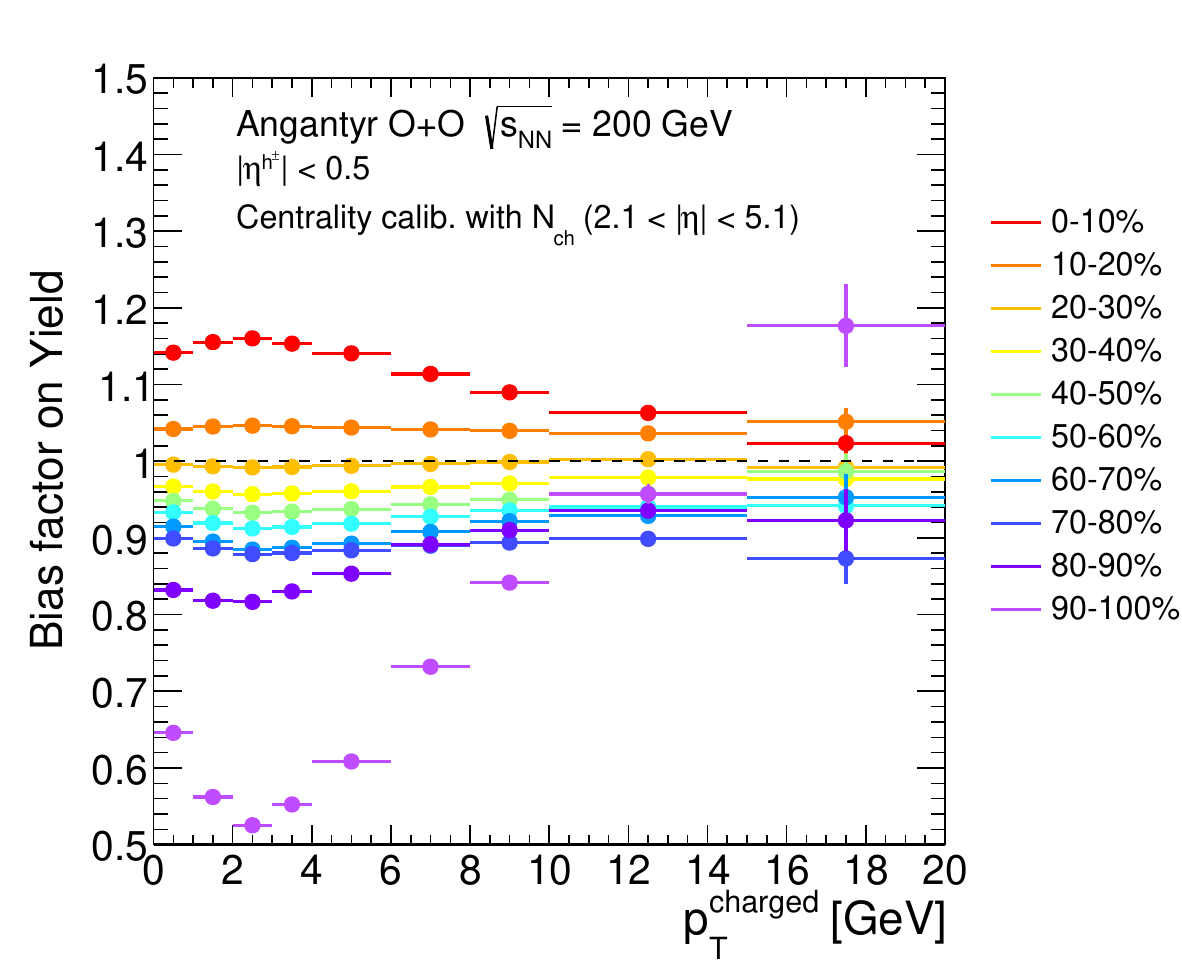}
\caption{\label{fig:RHICyieldsAngantyr} Centrality bias factors for charged-hadron yield measurements as a function of $p_\mathrm{T}$ in O+O collisions at RHIC, determined in \textsc{Angantyr}, using the $N_\mathrm{ch}$-based centrality definition. The different colors represent different selected centrality intervals.}
\end{figure}

\begin{figure}[!t]
\includegraphics[width=\linewidth]{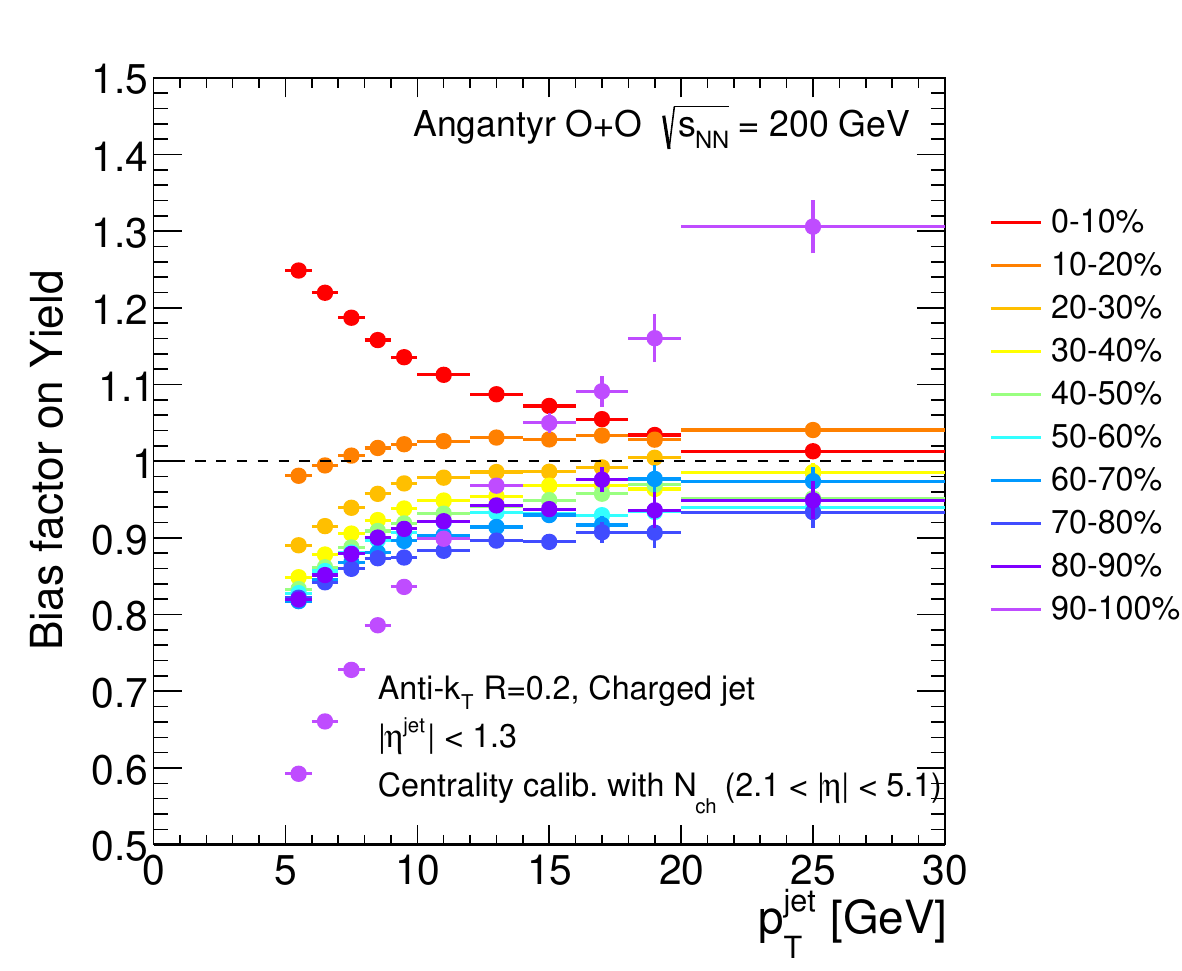}
\caption{\label{fig:RHICyieldsAngantyr2} Centrality bias factors for charged-jet yield measurements as a function of $p_\mathrm{T}$ in O+O collisions at RHIC, determined in \textsc{Angantyr}, using the $N_\mathrm{ch}$-based centrality definition. The different colors represent different selected centrality intervals.}
\end{figure}

We quantify bias effects in 10\%-wide centrality intervals for O+O collisions at RHIC and LHC over the full range 0--100\%. We explicitly include discussion of the problematic peripheral region, e.g. 80–-100\%, even though in practice it will likely not be experimentally accessible in O+O collisions. In Pb+Pb collisions, this region was already shown to be severely affected by selection bias effects~\cite{Loizides:2017sqq,ALICE:2018ekf}.

\section{O+O collisions at RHIC}

Fig.~\ref{fig:ncoll} shows example distributions of $N_\mathrm{coll}$ in centrality-selected O+O events at RHIC, without any hard process selection. Notably, the $N_\mathrm{coll}$ distributions in adjacent centrality bins have significant overlap. Thus, for an event at fixed $N_\mathrm{coll}$, the presence of a hard scattering that modifies particle production at forward rapidity may bias it towards a particular centrality bin. 

Figs.~\ref{fig:RHICyieldsAngantyr} and~\ref{fig:RHICyieldsAngantyr2} show the centrality bias factors for charged-hadron and charged-jet yields in O+O collisions at RHIC, determined with the \textsc{Angantyr} generator, using the $N_\mathrm{ch}$ centrality definition. At low $p_\mathrm{T}$, the bias factors for all but the most peripheral 90-100\% events are approximately in the range 0.8--1.2, with an increase in the most central events and then a centrality-differential ``ordering'' of decreasing bias factors in more peripheral events in which the sign of the bias reverses after 30\% centrality.
This modification pattern is broadly similar to that observed for $p$+A-type collisions in a similar $p_\mathrm{T}$ range~\cite{PHENIX:2013jxf,ALICE:2014xsp}, where it is attributed to the modest increase of soft particle production associated with the hard scattering. In the most peripheral events this effect leads to a significant jet veto, where imposing a low upper bound on the centrality signal strongly rejects events with hard-scattering processes. 

At significantly higher $p_\mathrm{T}$ in Figs.~\ref{fig:RHICyieldsAngantyr} and~\ref{fig:RHICyieldsAngantyr2} , the bias factors for all but the most peripheral events systematically tend towards unity. This suggests that the additional underlying event production associated with a hard scattering is systematically decreasing with increasing parton $p_\mathrm{T}$, and may eventually cause a ``reversal'' of the bias effect.
Such a change in behavior could be understood as a consequence of the initial-state scattered-parton kinematics. Producing a high-$p_\mathrm{T}$ parton which fragments into jets or hadrons at mid-rapidity requires the removal of two large-$x$ partons in the beam with $x \sim p_\mathrm{T} / E_\mathrm{beam}$. The remaining beam remnants which undergo soft interactions therefore do so at a lower effective $\sqrt{s}$ and produce a smaller centrality signal at forward rapidities~\cite{ATLAS:2015vql}. As the measured hadron or jet $p_\mathrm{T}$ increases, this effect progressively counteracts the initial multiplicity increase. 

Thus, the bias factors for charged particles may be thought of as being dominated by different physics effects in different $p_\mathrm{T}$ regions: soft physics processes for $p_\mathrm{T} \lesssim 2$~GeV, the hard--UE correlation for the region $p_\mathrm{T} \approx 2$--$10$~GeV and the energy conservation effect for $p_\mathrm{T} \gtrsim 10$ GeV. As expected, the bias factors are qualitatively similar between hadrons and jets in terms of sign, overall magnitude, and $p_\mathrm{T}$ dependence, accounting for the fact that the jets are composed of hadrons. In practice, measuring in wider centrality bins such as 0--30\% would have a reduced effect compared to the most central 0--10\%, albeit at the cost of a potentially weaker jet quenching signal.

In the experimental measurements, heavy-ion collision events generally require a minimum-bias (MB) trigger or offline selection requirement which captures a large but not complete fraction of the full inelastic cross-section, and is therefore not fully efficient in the most peripheral centrality bins. Thus, the recorded 90-100\% events are particularly subject to an additional MB trigger bias which selects for more active events, introducing experimental and modeling challenges. An additional complication is the strong sensitivity to the so-called anchor point used in the centrality analysis~\cite{Jonas:2021xju}. Explicitly modeling the MB trigger bias effect will depend on the details of the experimental setup and would require a full \textsc{GEANT}~\cite{GEANT4:2002zbu} simulation.

By comparing the bias factors in central and peripheral events in Figs.~\ref{fig:RHICyieldsAngantyr} and~\ref{fig:RHICyieldsAngantyr2}, we note that a measurement of the central-to-peripheral ratio, $R_\mathrm{CP}$, at RHIC energies would be significantly enhanced above unity at low-$p_\mathrm{T}$ but then would systematically decrease with increasing $p_\mathrm{T}$, all from centrality bias effects alone and before the inclusion of other physics. This decreasing trend is potentially compatible with the $p_\mathrm{T}$ dependence of preliminary measurements recently presented by STAR~\cite{Zhang:2025cgr}. 


\begin{figure}[!t]
\includegraphics[width=\linewidth]{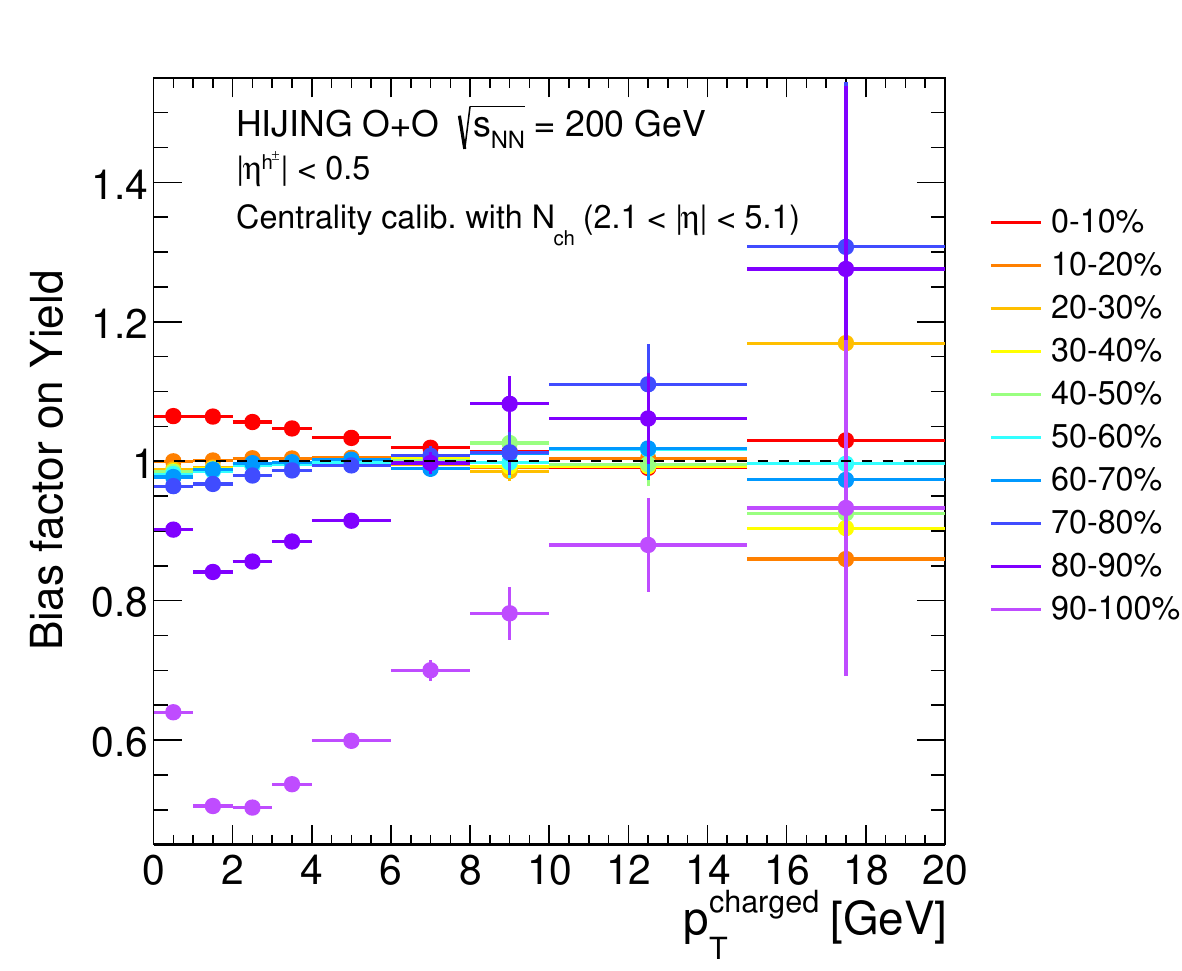}
\caption{\label{fig:RHICyieldsHijing} Centrality bias factors for charged-hadron yield measurements as a function of $p_\mathrm{T}$ in O+O collisions at RHIC, determined in \textsc{Hijing}, using the $N_\mathrm{ch}$-based centrality definition. The different colors represent different selected centrality intervals.}
\end{figure}

The analysis procedure was also repeated in the older \textsc{Hijing} event generator, with Fig.~\ref{fig:RHICyieldsHijing} showing an illustrative example of charged-hadron bias factors. The sign of the bias effect and the overall trends with $p_\mathrm{T}$ are similar. However, the magnitudes of the modifications are smaller, and are within the approximate range 0.85--1.05 in all but the most peripheral 90--100\% events. The difference between \textsc{Angantyr} and \textsc{Hijing} indicates the level of sensitivity to the choice of model, and highlights the need to have a good match to key aspects of the data (e.g., the forward $N_\mathrm{ch}$ or $\Sigma E_\mathrm{T}$ distributions) before applying bias factors as corrections to centrality-dependent measurements in data.

\begin{figure}[!t]
\includegraphics[width=\linewidth]{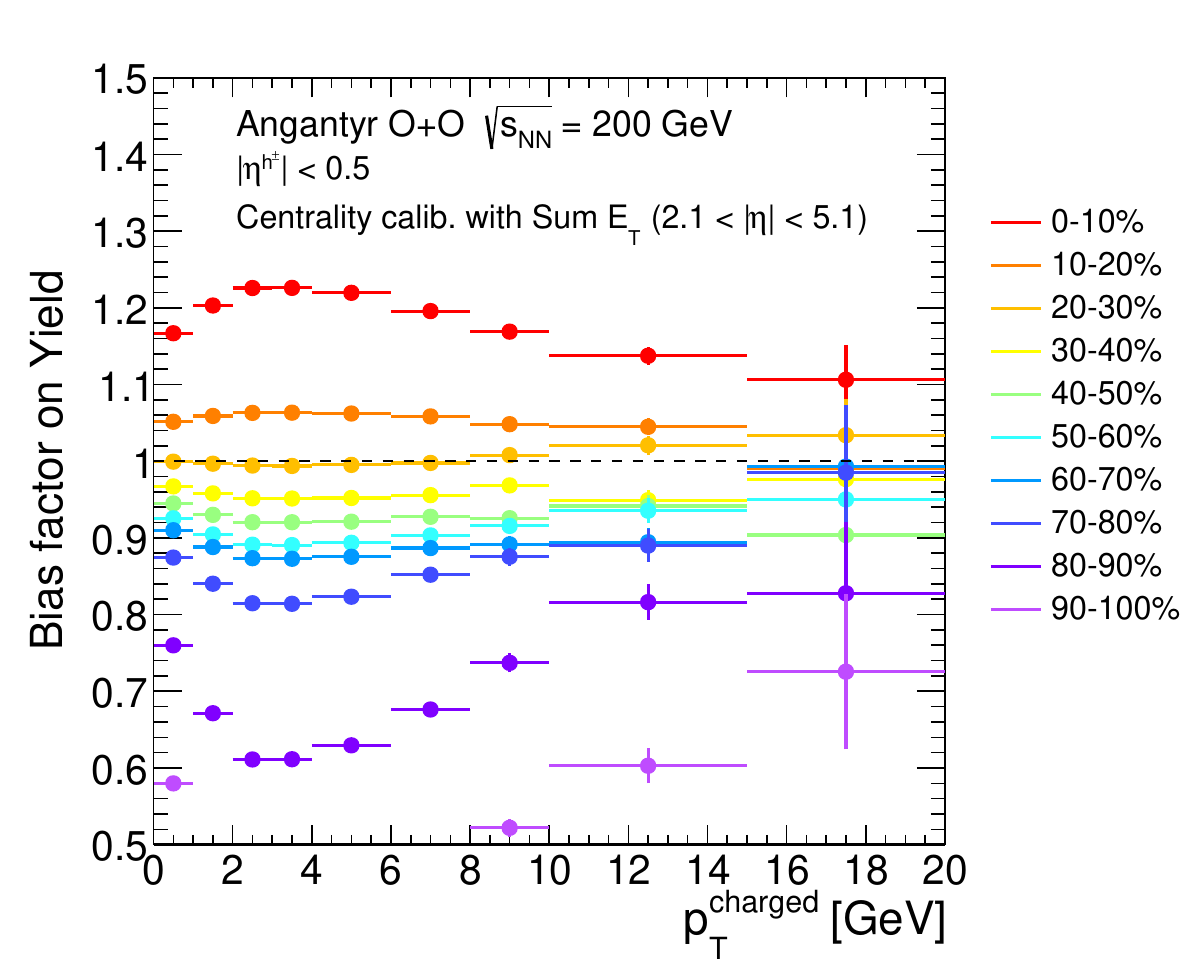}
\caption{\label{fig:RHICyieldsAngantyrSumET} Centrality bias factors for charged-hadron yield measurements as a function of $p_\mathrm{T}$ in O+O collisions at RHIC, determined in \textsc{Angantyr}, using the $\Sigma{E}_{T}$-based centrality definition. The different colors represent different selected centrality intervals.}
\end{figure}

To explore the sensitivity to $N_\mathrm{ch}$ vs. $\Sigma{E}_\mathrm{T}$-based definitions, Fig.~\ref{fig:RHICyieldsAngantyrSumET} shows the charged-hadron bias factors in \textsc{Angantyr}, but now using the $\Sigma{E}_\mathrm{T}$ centrality definition. Compared to Fig.~\ref{fig:RHICyieldsAngantyr}, the bias factors are moderately larger but have a qualitatively similar centrality ordering and $p_\mathrm{T}$ dependence. 

While the STAR EPD is principally counting the multiplicity $N_\mathrm{ch}$, it may also be sensitive to hadron showers which began upstream of the detector and thus have an effective contribution to the centrality signal from $\Sigma{E}_\mathrm{T}$. For this reason, again a full \textsc{GEANT} simulation would be necessary to determine the bias factors with precision.

\begin{figure}[!t]
\includegraphics[width=\linewidth]{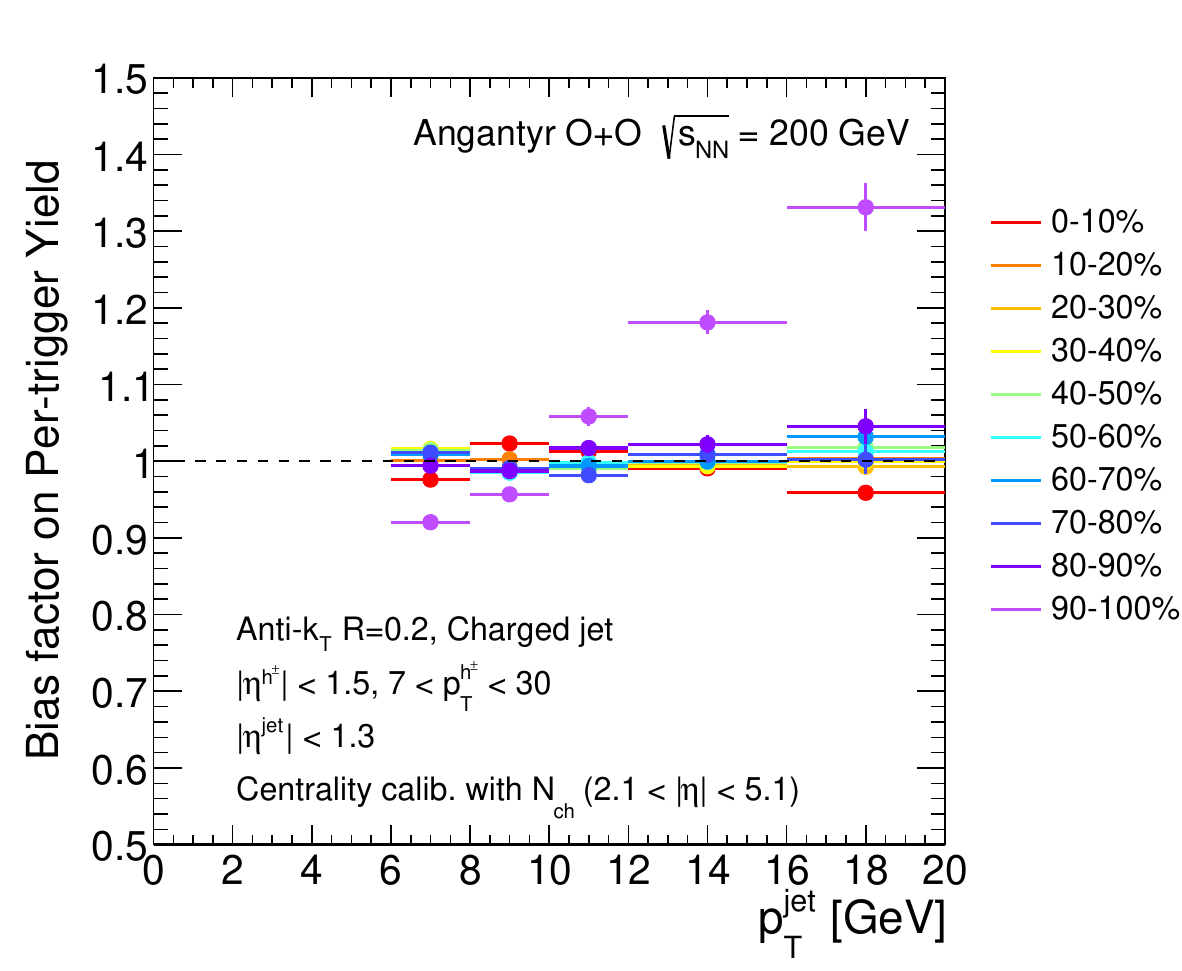}
\caption{\label{fig:RHICcorrAngantyr} Centrality bias factors for hadron-triggered semi-inclusive recoil jet yields, as a function of jet $p_\mathrm{T}$ in O+O collisions at RHIC, determined in \textsc{Angantyr}, using the $N_\mathrm{ch}$-based centrality definition. The different colors represent different selected centrality intervals.}
\end{figure}

\begin{figure}[!h]
\includegraphics[width=0.95\linewidth]{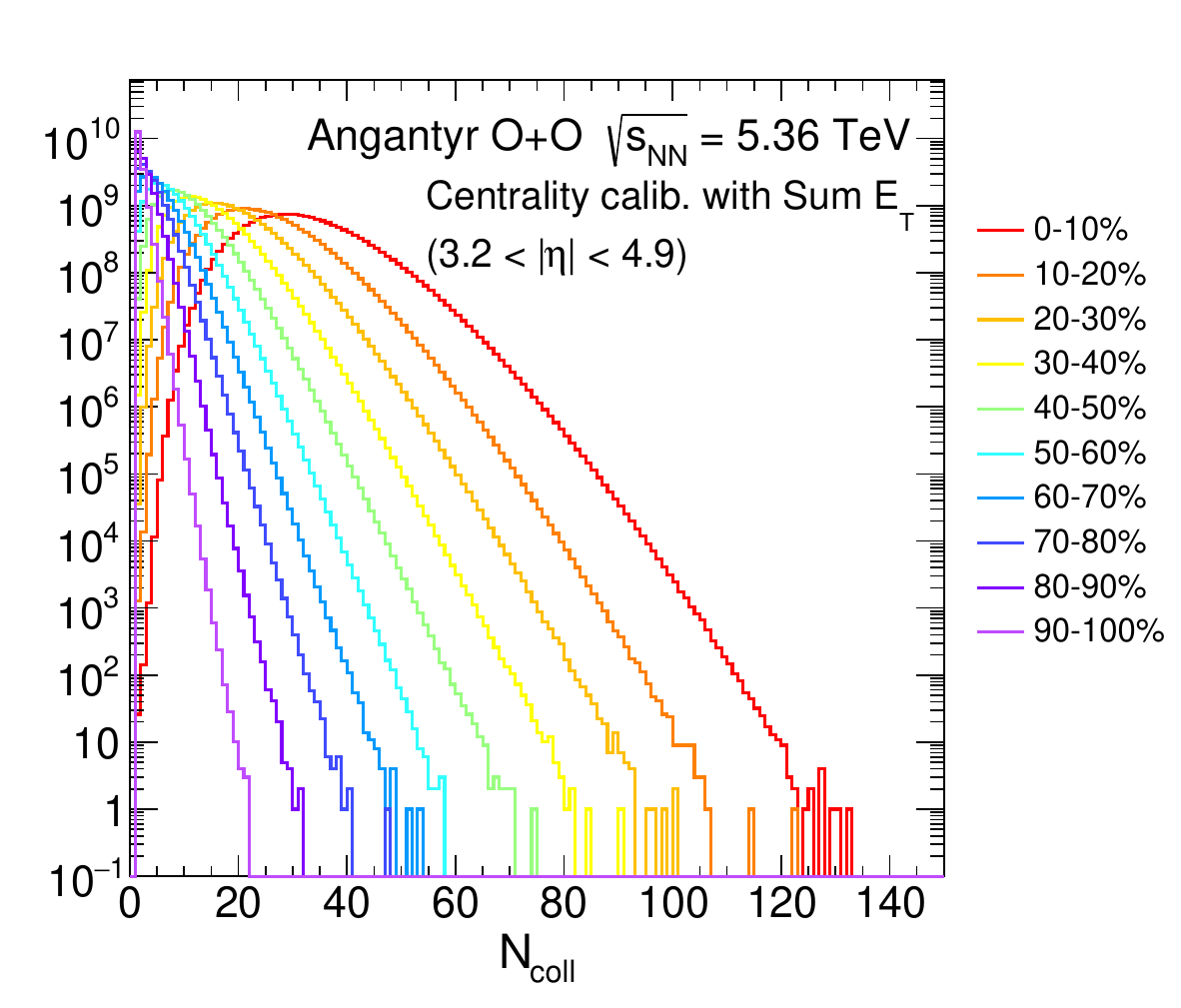}
\caption{\label{fig:ncollLHC} $N_\mathrm{coll}$ distributions in centrality-selected O+O events in the \textsc{Angantyr} heavy-ion event generator. Results are shown for LHC with a forward transverse energy sum definition, $\Sigma{E}_\mathrm{T}$.}
\end{figure}

In addition to single inclusive yields, measurements of intra-event correlations such as di-jet, hadron+jet, and $\gamma$+jet are widely used as probes of QGP jet quenching effects. As an illustrative example, Fig.~\ref{fig:RHICcorrAngantyr} shows the bias factors for hadron-triggered associated jet yields as a function of jet $p_\mathrm{T}$. Strikingly in comparison to the inclusive yields above, for all but the most peripheral 90-100\% events, all bias factors are a few percent away from unity, similar to the statistical uncertainties in the MC event sample. This much decreased sensitivity to the centrality bias comes from the nature of the per-trigger semi-inclusive observable, where the bias applies to both trigger and associated yields to approximately the same degree. For the most peripheral 90--100\% bin, the behavior may arise from an effective veto on additional jet production at forward rapidities, thus forcing any hadron--triggered events to have only a single associated jet at mid-rapidity which captures more of the balancing parton's $p_\mathrm{T}$. 

We note that in Ref.~\cite{ALICE:2023plt}, a very high-multiplicity requirement at forward rapidities in $p$+$p$ collisions at the LHC was found to select imbalanced di-jet pairs at mid-rapidity. This may arise from a ``direct'' bias where the associated jet is at more forward rapidities and itself contributes to the centrality signal, thus decreasing the per-jet associated yield at mid-rapidity. Such a bias must surely be present for correlation measurements in O+O collisions as well, but is apparently negligible within the broad centrality selection of 0--10\% and the much larger degree of soft particle production. This suggests that coincidence measurements in O+O collisions such as hadron--jet, di-jet, or $\gamma$+jet correlations may be particularly fruitful ways to search for jet quenching signatures in the most central events.

\section{O+O collisions at the LHC}

\begin{figure}[!t]
\includegraphics[width=\linewidth]{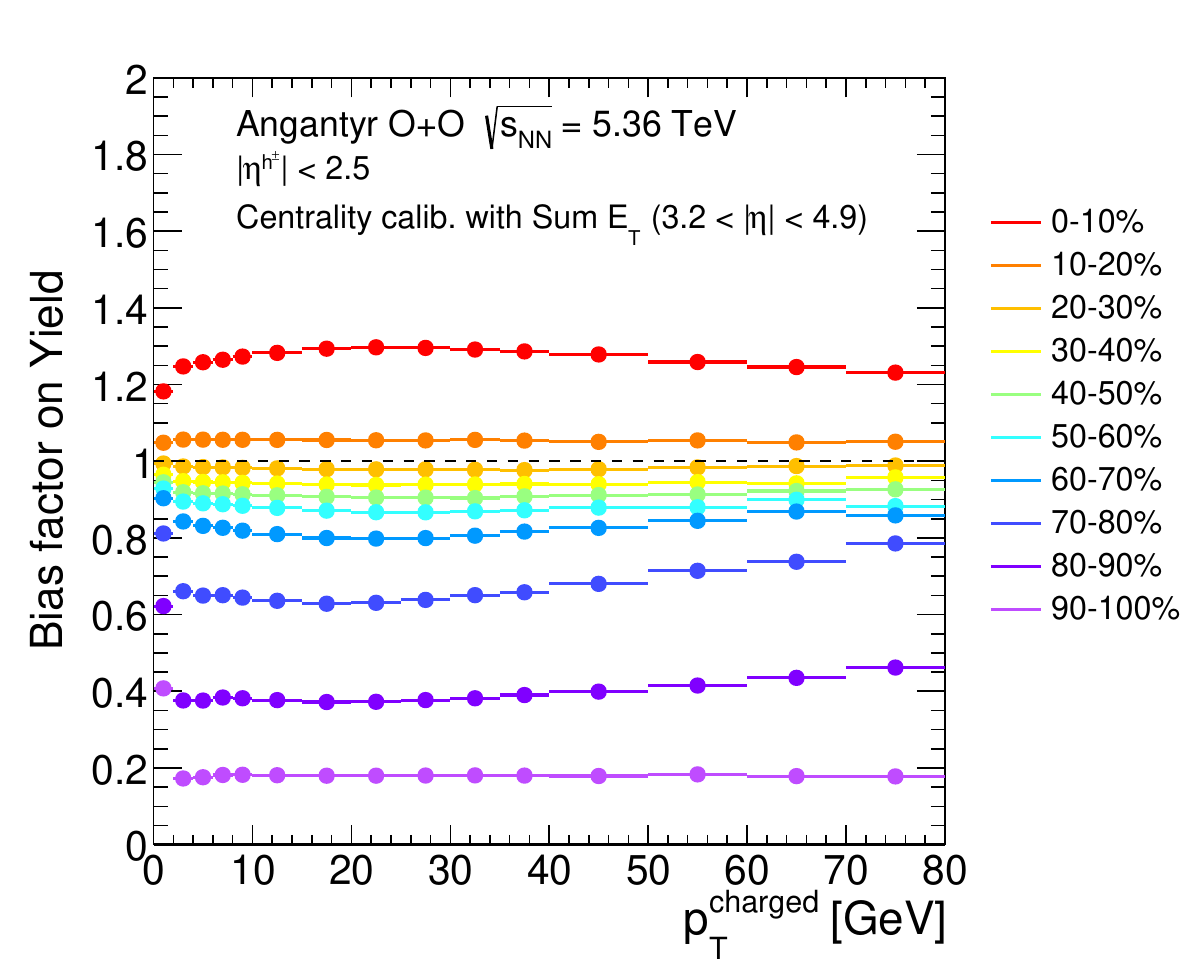}
\caption{\label{fig:LHCyieldsAngantyr} Centrality bias factors for charged-hadron yield measurements as a function of $p_\mathrm{T}$ in O+O collisions at LHC, determined in \textsc{Angantyr}, for the $\Sigma E_\mathrm{T}$ centrality definition. The different colors represent different selected centrality intervals.}
\end{figure}

\begin{figure}[!t]
\includegraphics[width=\linewidth]{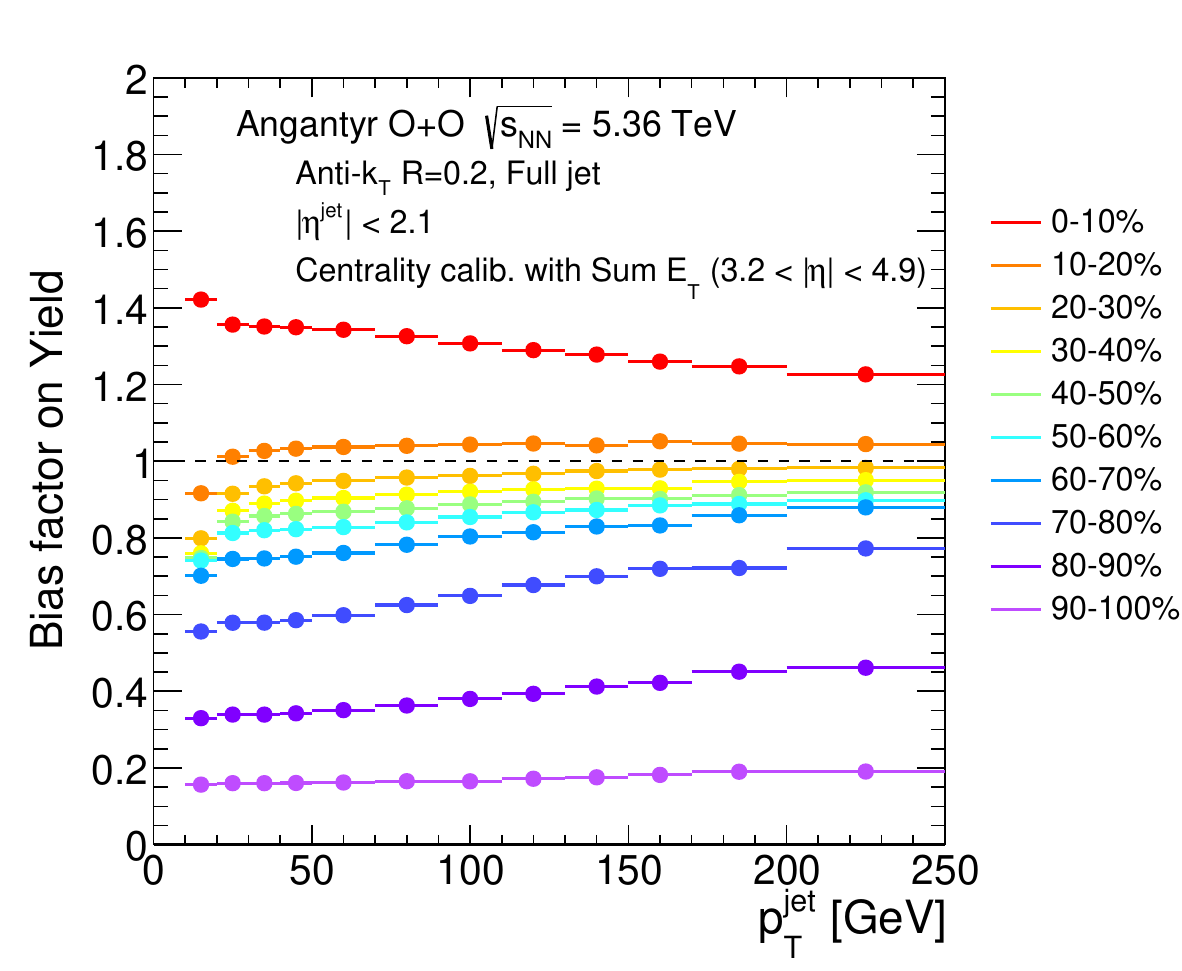}
\caption{\label{fig:LHCyieldsAngantyr2} Centrality bias factors for full-jet yield measurements as a function of $p_\mathrm{T}$ in O+O collisions at LHC, determined in \textsc{Angantyr}, for the $\Sigma E_\mathrm{T}$ centrality definition. The different colors represent different selected centrality intervals.}
\end{figure}

Fig.~\ref{fig:ncollLHC} shows example distributions of $N_\mathrm{coll}$ in centrality-selected O+O events at the LHC, without any hard process selection, where the centrality definition is based on the forward transverse energy sum, $\Sigma E_\mathrm{T}$. As in Fig.~\ref{fig:ncoll} for RHIC energies above, the $N_\mathrm{coll}$ distributions in adjacent centrality bins have significant overlap, again suggesting the presence of bias effects for hard-process measurements.

Figs.~\ref{fig:LHCyieldsAngantyr} and~\ref{fig:LHCyieldsAngantyr2} show the centrality bias factors for charged-hadron and full jet yields in O+O collisions at the LHC, determined with the \textsc{Angantyr} generator, for the $\Sigma{E}_\mathrm{T}$ centrality definition. The bias effects are significantly stronger than the analogous ones at RHIC in Figs.~\ref{fig:RHICyieldsAngantyr} and~\ref{fig:RHICyieldsAngantyr2}. In the most central events, they enhance the yield above unity by approximately 1.3, and they are ordered in a way which is systematic in centrality, with bias factors smaller than 0.8 for all selections more peripheral than 60\%. A significantly larger bias at LHC compared to RHIC energies was also found for $p$+A-type collisions in Ref.~\cite{PHENIX:2013jxf}.

For the $p_\mathrm{T}$ ranges considered here, the bias factors have a much weaker $p_\mathrm{T}$ dependence and, over the shown kinematic range, do not convergence to unity as seen in the RHIC case in Figs.~\ref{fig:RHICyieldsAngantyr} and~\ref{fig:RHICyieldsAngantyr2}. Above, this behavior was attributed to the depletion of the beam-remnant energy from removing a large-$x$ parton via the hard scattering. In the LHC case, the expected luminosities in O+O are not sufficient to reach this regime, since jet production at $p_\mathrm{T} \sim 100$~GeV typically arises from typical parton configurations with only $x \sim p_\mathrm{T}^\mathrm{jet} / E_\mathrm{beam} \sim 0.05$. 

\begin{figure}[!t]
\includegraphics[width=\linewidth]{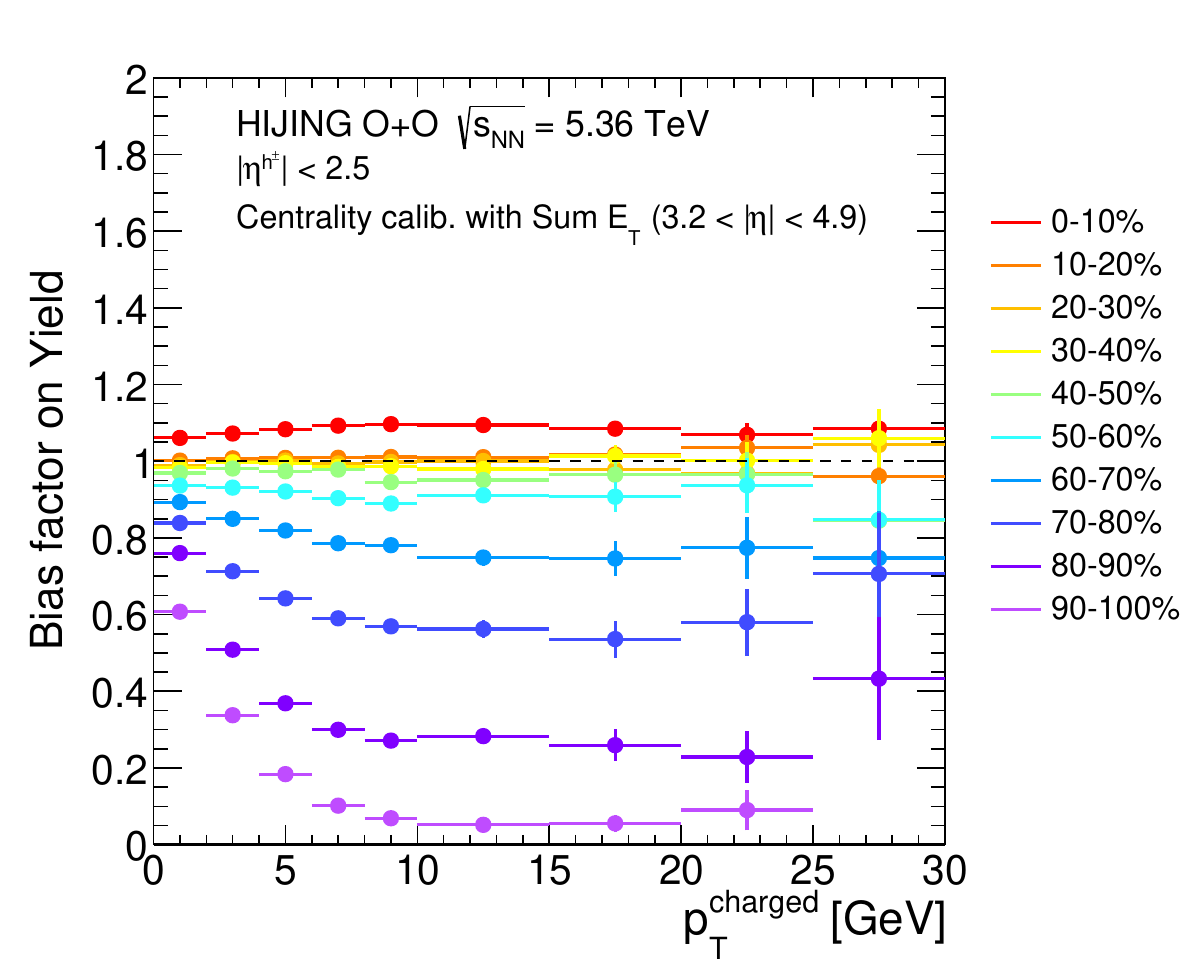}
\caption{\label{fig:LHCyieldsHijing} Centrality bias factors for charged-hadron yield measurements as a function of $p_\mathrm{T}$ in O+O collisions at LHC, determined in \textsc{Hijing}, for the $\Sigma E_\mathrm{T}$ centrality definition. The different colors represent different selected centrality intervals.}
\end{figure}

\begin{figure}[!t]
\includegraphics[width=\linewidth]{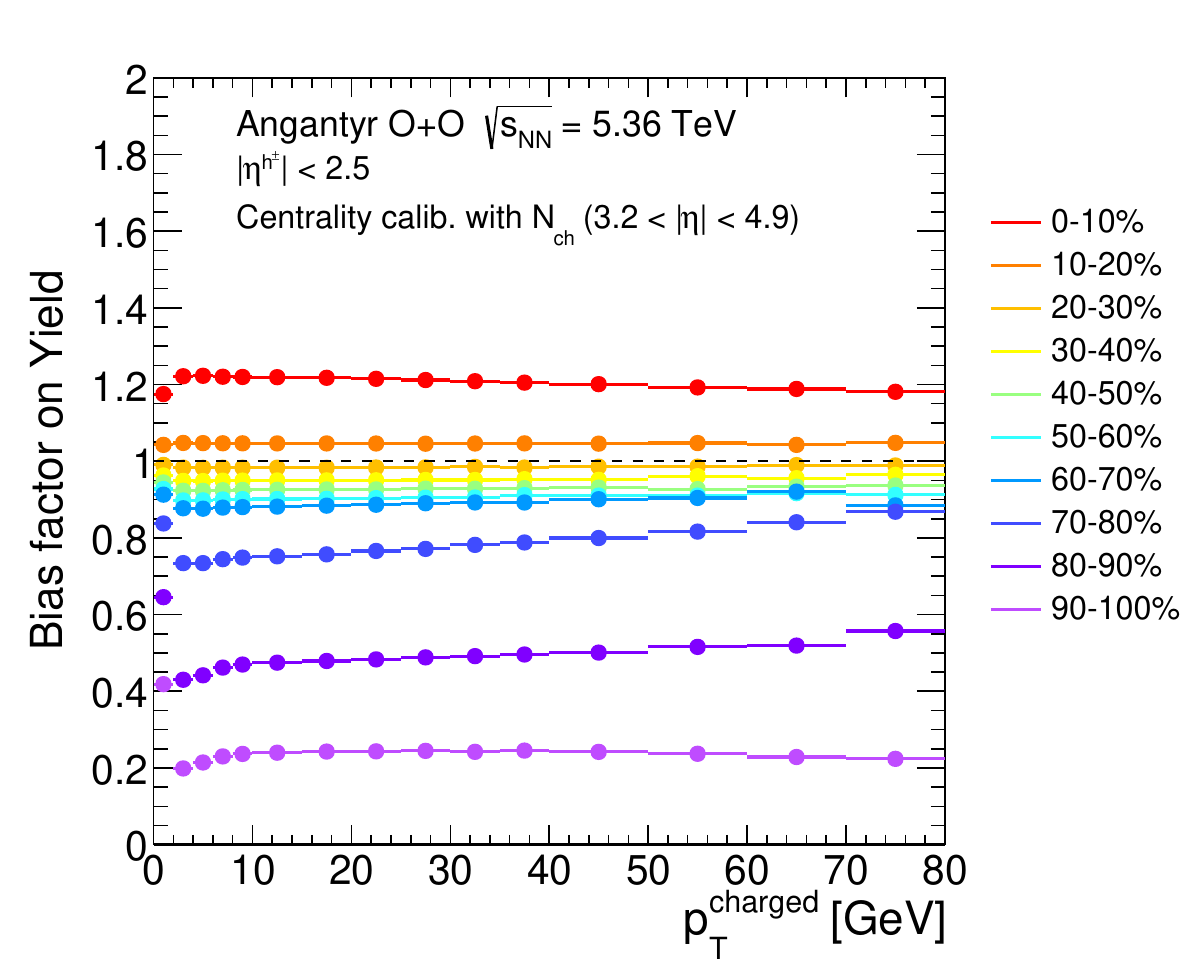}
\caption{\label{fig:LHCyieldsAngantyrNch} Centrality bias factors for charged-hadron yield measurements as a function of $p_\mathrm{T}$ in O+O collisions at LHC, determined in \textsc{Angantyr}, for the $N_\mathrm{ch}$ centrality definition. The different colors represent different selected centrality intervals.}
\end{figure}

Fig.~\ref{fig:LHCyieldsHijing} shows the analogous bias factors for charged hadrons in Fig.~\ref{fig:LHCyieldsAngantyr},  determined in the \textsc{Hijing} generator, over a narrower $p_\mathrm{T}$ range given the smaller simulated sample. As seen in the RHIC case above, the qualitative behavior is similar to \textsc{Angantyr}, albeit with different magnitudes of the bias effects. For example, the bias factors for events in the 0-40\% centrality range are all individually much closer to unity.

To check the sensitivity to centrality definition,  Fig.~\ref{fig:LHCyieldsAngantyrNch} shows the analogous set of bias factors for charged particles at the LHC but for the $N_\mathrm{ch}$ centrality definition at forward rapidity. As with the case at RHIC energies, the bias effects are moderately smaller but have qualitatively similar trends, again highlighting the need for a detailed simulation of experimental effects for a precision determination of the bias factors. 


\begin{figure}[!h]
\includegraphics[width=\linewidth]{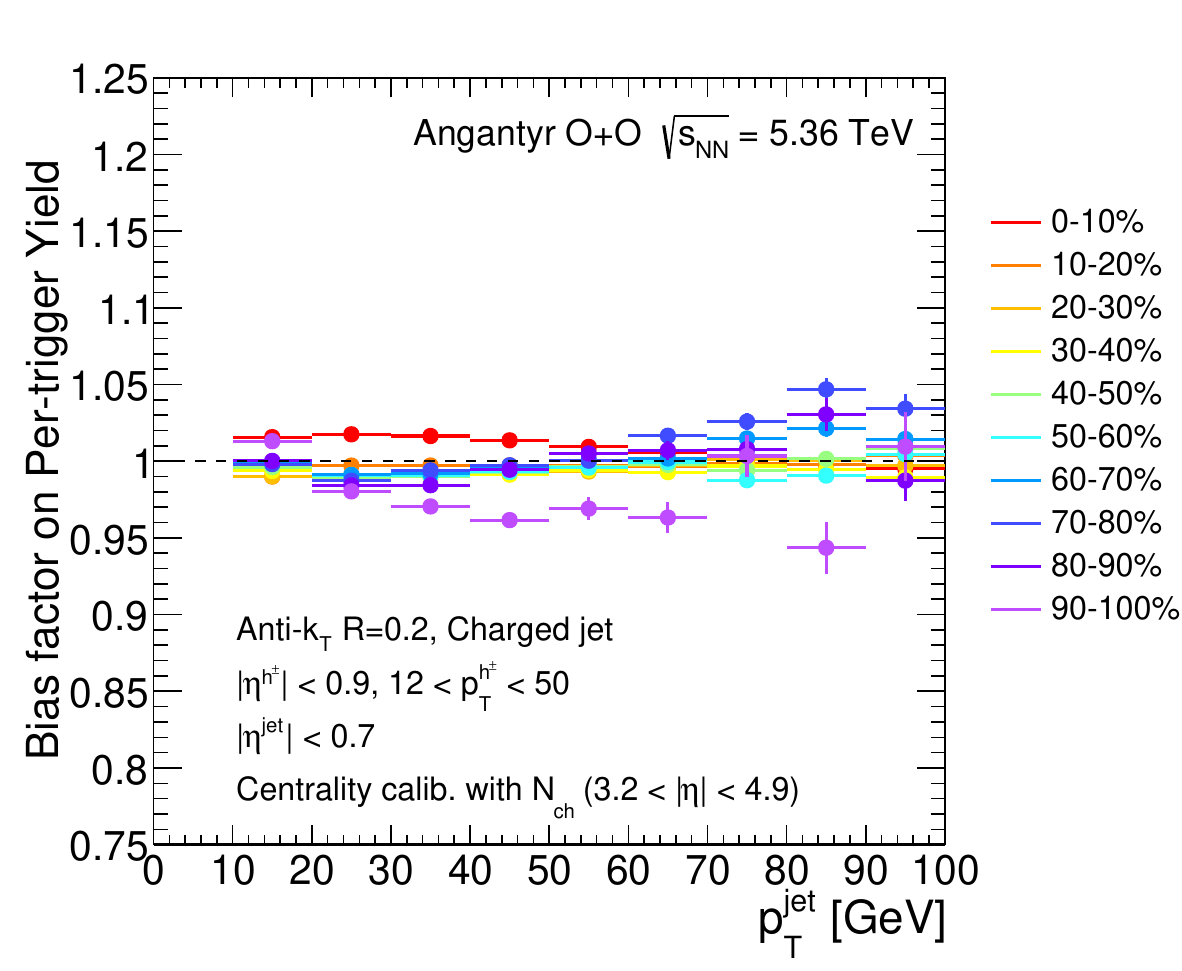}
\caption{\label{fig:LHCcorrAngantyr} Centrality bias factors for hadron-triggered semi-inclusive recoil jet yields, as a function of jet $p_\mathrm{T}$ in O+O collisions at LHC, determined in \textsc{Angantyr}, using the $N_\mathrm{ch}$ centrality definition. The different colors represent different selected centrality intervals.}
\end{figure}

Fig.~\ref{fig:LHCcorrAngantyr} shows the bias factors for an illustrative correlation measurement, which is the per-hadron-triggered associated jet yield, using the $N_\mathrm{ch}$ centrality definition. Similar to Fig.~\ref{fig:RHICcorrAngantyr} at RHIC energies, the bias factors for correlation measurements at the LHC are much closer to unity than they are for inclusive yield measurements, resulting in only small deviations. This again highlights the relative resilience of correlation measurements compared to yield measurements, and suggests they may be particularly fruitful ways to search for jet quenching signatures in the most central O+O collisions.

For all of the measurements discussed above, the rapidity separation between the hard-probe measurement region and the centrality-defining region is also expected to play a role, with a larger sensitivity to centrality bias effects for smaller separations (i.e. in LHC detectors with a large pseudorapidity acceptance in the central region). Finally, the presence of true jet quenching effects is expected to modestly lessen the magnitude of centrality bias effects, since this would result in hard process yields that grow more slowly with centrality than expected from the $N_\mathrm{coll}$ scaling assumption used here.

\section{Conclusion}

Oxygen+Oxygen collisions at RHIC and LHC offer a compelling way to search for jet quenching or other modifications of hard processes in intermediate-sized nuclear collision systems. Since the magnitude of jet quenching effects in O+O events is expected to be small compared to that in central nucleus--nucleus collisions, it is important to account for confounding effects on commonly used observables. We examine the well-known multiplicity selection bias, in which the yields in centrality-selected events may be enhanced or suppressed compared to the true yield for events of the given geometry class, due to a correlation between the hard process and the centrality signal. Our study examines two Monte Carlo event generators and a selection of centrality-definition regions and final-state observables similar to those previously used by experiments at RHIC and LHC. Centrality bias effects on measurements of hadron and jet yields, and thus on nuclear modification factors, at RHIC and the LHC can reach 20--30\% for central to mid-peripheral events, with a dependence on the collider energy, kinematics, centrality definition, and generator. On the other hand, they are essentially absent for correlation measurements such hadron-triggered jet yields. The studies in this paper may inform the design of experimental measurements, such as the choice of centrality definition (multiplicity- vs. transverse-energy-based) or observable (yields vs. correlations), and aid in the interpretation of experimental results.

\begin{acknowledgments}
The authors are grateful to Anthony Timmins for discussion. The authors acknowledge support by the DOE Office of Science grants DE-FG02-03ER41244 (JP, JLN,  DVP) and DE-SC0005131 (CL), and by the National Research Foundation of Korea (NRF) grant funded by the Korea government (MSIT) under Contract No. RS-2025-00554431 (S.L.). 
\end{acknowledgments}

\bibliography{apssamp}

\begin{thebibliography}{53}%
\makeatletter
\providecommand \@ifxundefined [1]{%
 \@ifx{#1\undefined}
}%
\providecommand \@ifnum [1]{%
 \ifnum #1\expandafter \@firstoftwo
 \else \expandafter \@secondoftwo
 \fi
}%
\providecommand \@ifx [1]{%
 \ifx #1\expandafter \@firstoftwo
 \else \expandafter \@secondoftwo
 \fi
}%
\providecommand \natexlab [1]{#1}%
\providecommand \enquote  [1]{``#1''}%
\providecommand \bibnamefont  [1]{#1}%
\providecommand \bibfnamefont [1]{#1}%
\providecommand \citenamefont [1]{#1}%
\providecommand \href@noop [0]{\@secondoftwo}%
\providecommand \href [0]{\begingroup \@sanitize@url \@href}%
\providecommand \@href[1]{\@@startlink{#1}\@@href}%
\providecommand \@@href[1]{\endgroup#1\@@endlink}%
\providecommand \@sanitize@url [0]{\catcode `\\12\catcode `\$12\catcode
  `\&12\catcode `\#12\catcode `\^12\catcode `\_12\catcode `\%12\relax}%
\providecommand \@@startlink[1]{}%
\providecommand \@@endlink[0]{}%
\providecommand \url  [0]{\begingroup\@sanitize@url \@url }%
\providecommand \@url [1]{\endgroup\@href {#1}{\urlprefix }}%
\providecommand \urlprefix  [0]{URL }%
\providecommand \Eprint [0]{\href }%
\providecommand \doibase [0]{https://doi.org/}%
\providecommand \selectlanguage [0]{\@gobble}%
\providecommand \bibinfo  [0]{\@secondoftwo}%
\providecommand \bibfield  [0]{\@secondoftwo}%
\providecommand \translation [1]{[#1]}%
\providecommand \BibitemOpen [0]{}%
\providecommand \bibitemStop [0]{}%
\providecommand \bibitemNoStop [0]{.\EOS\space}%
\providecommand \EOS [0]{\spacefactor3000\relax}%
\providecommand \BibitemShut  [1]{\csname bibitem#1\endcsname}%
\let\auto@bib@innerbib\@empty
\bibitem [{\citenamefont {Brewer}\ \emph {et~al.}(2021)\citenamefont {Brewer},
  \citenamefont {Mazeliauskas},\ and\ \citenamefont {van~der
  Schee}}]{Brewer:2021kiv}%
  \BibitemOpen
  \bibfield  {author} {\bibinfo {author} {\bibfnamefont {J.}~\bibnamefont
  {Brewer}}, \bibinfo {author} {\bibfnamefont {A.}~\bibnamefont
  {Mazeliauskas}},\ and\ \bibinfo {author} {\bibfnamefont {W.}~\bibnamefont
  {van~der Schee}},\ }\bibfield  {title} {\bibinfo {title} {{Opportunities of
  OO and $p$O collisions at the LHC}},\ }in\ \href@noop {} {\emph {\bibinfo
  {booktitle} {{Opportunities of OO and pO collisions at the LHC}}}}\ (\bibinfo
  {year} {2021})\ \Eprint {https://arxiv.org/abs/2103.01939} {arXiv:2103.01939
  [hep-ph]} \BibitemShut {NoStop}%
\bibitem [{\citenamefont {Cunqueiro}\ and\ \citenamefont
  {Sickles}(2022)}]{Cunqueiro:2021wls}%
  \BibitemOpen
  \bibfield  {author} {\bibinfo {author} {\bibfnamefont {L.}~\bibnamefont
  {Cunqueiro}}\ and\ \bibinfo {author} {\bibfnamefont {A.~M.}\ \bibnamefont
  {Sickles}},\ }\bibfield  {title} {\bibinfo {title} {{Studying the QGP with
  Jets at the LHC and RHIC}},\ }\href
  {https://doi.org/10.1016/j.ppnp.2022.103940} {\bibfield  {journal} {\bibinfo
  {journal} {Prog. Part. Nucl. Phys.}\ }\textbf {\bibinfo {volume} {124}},\
  \bibinfo {pages} {103940} (\bibinfo {year} {2022})},\ \Eprint
  {https://arxiv.org/abs/2110.14490} {arXiv:2110.14490 [nucl-ex]} \BibitemShut
  {NoStop}%
\bibitem [{\citenamefont {Loizides}(2016)}]{Loizides:2016tew}%
  \BibitemOpen
  \bibfield  {author} {\bibinfo {author} {\bibfnamefont {C.}~\bibnamefont
  {Loizides}},\ }\bibfield  {title} {\bibinfo {title} {{Experimental overview
  on small collision systems at the LHC}},\ }\href
  {https://doi.org/10.1016/j.nuclphysa.2016.04.022} {\bibfield  {journal}
  {\bibinfo  {journal} {Nucl. Phys. A}\ }\textbf {\bibinfo {volume} {956}},\
  \bibinfo {pages} {200} (\bibinfo {year} {2016})},\ \Eprint
  {https://arxiv.org/abs/1602.09138} {arXiv:1602.09138 [nucl-ex]} \BibitemShut
  {NoStop}%
\bibitem [{\citenamefont {Nagle}\ and\ \citenamefont
  {Zajc}(2018)}]{Nagle:2018nvi}%
  \BibitemOpen
  \bibfield  {author} {\bibinfo {author} {\bibfnamefont {J.~L.}\ \bibnamefont
  {Nagle}}\ and\ \bibinfo {author} {\bibfnamefont {W.~A.}\ \bibnamefont
  {Zajc}},\ }\bibfield  {title} {\bibinfo {title} {{Small System Collectivity
  in Relativistic Hadronic and Nuclear Collisions}},\ }\href
  {https://doi.org/10.1146/annurev-nucl-101916-123209} {\bibfield  {journal}
  {\bibinfo  {journal} {Ann. Rev. Nucl. Part. Sci.}\ }\textbf {\bibinfo
  {volume} {68}},\ \bibinfo {pages} {211} (\bibinfo {year} {2018})},\ \Eprint
  {https://arxiv.org/abs/1801.03477} {arXiv:1801.03477 [nucl-ex]} \BibitemShut
  {NoStop}%
\bibitem [{\citenamefont {Acharya}\ \emph {et~al.}(2018)\citenamefont {Acharya}
  \emph {et~al.}}]{ALICE:2017svf}%
  \BibitemOpen
  \bibfield  {author} {\bibinfo {author} {\bibfnamefont {S.}~\bibnamefont
  {Acharya}} \emph {et~al.} (\bibinfo {collaboration} {ALICE}),\ }\bibfield
  {title} {\bibinfo {title} {{Constraints on jet quenching in p-Pb collisions
  at $\mathbf{\sqrt{s_{NN}}}$ = 5.02 TeV measured by the event-activity
  dependence of semi-inclusive hadron-jet distributions}},\ }\href
  {https://doi.org/10.1016/j.physletb.2018.05.059} {\bibfield  {journal}
  {\bibinfo  {journal} {Phys. Lett. B}\ }\textbf {\bibinfo {volume} {783}},\
  \bibinfo {pages} {95} (\bibinfo {year} {2018})},\ \Eprint
  {https://arxiv.org/abs/1712.05603} {arXiv:1712.05603 [nucl-ex]} \BibitemShut
  {NoStop}%
\bibitem [{\citenamefont {Aad}\ \emph {et~al.}(2023{\natexlab{a}})\citenamefont
  {Aad} \emph {et~al.}}]{ATLAS:2022iyq}%
  \BibitemOpen
  \bibfield  {author} {\bibinfo {author} {\bibfnamefont {G.}~\bibnamefont
  {Aad}} \emph {et~al.} (\bibinfo {collaboration} {ATLAS}),\ }\bibfield
  {title} {\bibinfo {title} {{Strong Constraints on Jet Quenching in
  Centrality-Dependent p+Pb Collisions at 5.02~TeV from ATLAS}},\ }\href
  {https://doi.org/10.1103/PhysRevLett.131.072301} {\bibfield  {journal}
  {\bibinfo  {journal} {Phys. Rev. Lett.}\ }\textbf {\bibinfo {volume} {131}},\
  \bibinfo {pages} {072301} (\bibinfo {year} {2023}{\natexlab{a}})},\ \Eprint
  {https://arxiv.org/abs/2206.01138} {arXiv:2206.01138 [nucl-ex]} \BibitemShut
  {NoStop}%
\bibitem [{\citenamefont {Chekhovsky}\ \emph {et~al.}(2025)\citenamefont
  {Chekhovsky} \emph {et~al.}}]{CMS:2025jbv}%
  \BibitemOpen
  \bibfield  {author} {\bibinfo {author} {\bibfnamefont {V.}~\bibnamefont
  {Chekhovsky}} \emph {et~al.} (\bibinfo {collaboration} {CMS}),\ }\bibfield
  {title} {\bibinfo {title} {{Search for jet quenching with dijets from
  high-multiplicity pPb collisions at $\sqrt{{s}_{\text{NN}}}$ = 8.16 TeV}},\
  }\href {https://doi.org/10.1007/JHEP07(2025)118} {\bibfield  {journal}
  {\bibinfo  {journal} {JHEP}\ }\textbf {\bibinfo {volume} {07}},\ \bibinfo
  {pages} {118}},\ \Eprint {https://arxiv.org/abs/2504.08507} {arXiv:2504.08507
  [nucl-ex]} \BibitemShut {NoStop}%
\bibitem [{\citenamefont {Aad}\ \emph {et~al.}(2024)\citenamefont {Aad} \emph
  {et~al.}}]{ATLAS:2023zfx}%
  \BibitemOpen
  \bibfield  {author} {\bibinfo {author} {\bibfnamefont {G.}~\bibnamefont
  {Aad}} \emph {et~al.} (\bibinfo {collaboration} {ATLAS}),\ }\bibfield
  {title} {\bibinfo {title} {{Measurement of the Centrality Dependence of the
  Dijet Yield in p+Pb Collisions at sNN=8.16\,\,TeV with the ATLAS Detector}},\
  }\href {https://doi.org/10.1103/PhysRevLett.132.102301} {\bibfield  {journal}
  {\bibinfo  {journal} {Phys. Rev. Lett.}\ }\textbf {\bibinfo {volume} {132}},\
  \bibinfo {pages} {102301} (\bibinfo {year} {2024})},\ \Eprint
  {https://arxiv.org/abs/2309.00033} {arXiv:2309.00033 [nucl-ex]} \BibitemShut
  {NoStop}%
\bibitem [{\citenamefont {Abdulhamid}\ \emph {et~al.}(2024)\citenamefont
  {Abdulhamid} \emph {et~al.}}]{STAR:2024nwm}%
  \BibitemOpen
  \bibfield  {author} {\bibinfo {author} {\bibfnamefont {M.}~\bibnamefont
  {Abdulhamid}} \emph {et~al.} (\bibinfo {collaboration} {STAR}),\ }\bibfield
  {title} {\bibinfo {title} {{Correlations of event activity with hard and soft
  processes in p+Au collisions at sNN=200 GeV at the RHIC STAR experiment}},\
  }\href {https://doi.org/10.1103/PhysRevC.110.044908} {\bibfield  {journal}
  {\bibinfo  {journal} {Phys. Rev. C}\ }\textbf {\bibinfo {volume} {110}},\
  \bibinfo {pages} {044908} (\bibinfo {year} {2024})},\ \Eprint
  {https://arxiv.org/abs/2404.08784} {arXiv:2404.08784 [nucl-ex]} \BibitemShut
  {NoStop}%
\bibitem [{\citenamefont {Perepelitsa}(2024)}]{Perepelitsa:2024eik}%
  \BibitemOpen
  \bibfield  {author} {\bibinfo {author} {\bibfnamefont {D.~V.}\ \bibnamefont
  {Perepelitsa}},\ }\bibfield  {title} {\bibinfo {title} {{Contribution to
  differential $\pi^0$ and $\gamma_\mathrm{dir}$ modification in small systems
  from color fluctuation effects}},\ }\href
  {https://doi.org/10.1103/PhysRevC.110.L011901} {\bibfield  {journal}
  {\bibinfo  {journal} {Phys. Rev. C}\ }\textbf {\bibinfo {volume} {110}},\
  \bibinfo {pages} {L011901} (\bibinfo {year} {2024})},\ \Eprint
  {https://arxiv.org/abs/2404.17660} {arXiv:2404.17660 [nucl-th]} \BibitemShut
  {NoStop}%
\bibitem [{\citenamefont {Katz}\ \emph {et~al.}(2020)\citenamefont {Katz},
  \citenamefont {Prado}, \citenamefont {Noronha-Hostler},\ and\ \citenamefont
  {Suaide}}]{Katz:2019qwv}%
  \BibitemOpen
  \bibfield  {author} {\bibinfo {author} {\bibfnamefont {R.}~\bibnamefont
  {Katz}}, \bibinfo {author} {\bibfnamefont {C.~A.~G.}\ \bibnamefont {Prado}},
  \bibinfo {author} {\bibfnamefont {J.}~\bibnamefont {Noronha-Hostler}},\ and\
  \bibinfo {author} {\bibfnamefont {A.~A.~P.}\ \bibnamefont {Suaide}},\
  }\bibfield  {title} {\bibinfo {title} {{System-size scan of $D$ meson
  $R_{AA}$ and $v_n$ using PbPb, XeXe, ArAr, and OO collisions at energies
  available at the CERN Large Hadron Collider}},\ }\href
  {https://doi.org/10.1103/PhysRevC.102.041901} {\bibfield  {journal} {\bibinfo
   {journal} {Phys. Rev. C}\ }\textbf {\bibinfo {volume} {102}},\ \bibinfo
  {pages} {041901} (\bibinfo {year} {2020})},\ \Eprint
  {https://arxiv.org/abs/1907.03308} {arXiv:1907.03308 [nucl-th]} \BibitemShut
  {NoStop}%
\bibitem [{\citenamefont {Huss}\ \emph
  {et~al.}(2021{\natexlab{a}})\citenamefont {Huss}, \citenamefont {Kurkela},
  \citenamefont {Mazeliauskas}, \citenamefont {Paatelainen}, \citenamefont
  {van~der Schee},\ and\ \citenamefont {Wiedemann}}]{Huss:2020dwe}%
  \BibitemOpen
  \bibfield  {author} {\bibinfo {author} {\bibfnamefont {A.}~\bibnamefont
  {Huss}}, \bibinfo {author} {\bibfnamefont {A.}~\bibnamefont {Kurkela}},
  \bibinfo {author} {\bibfnamefont {A.}~\bibnamefont {Mazeliauskas}}, \bibinfo
  {author} {\bibfnamefont {R.}~\bibnamefont {Paatelainen}}, \bibinfo {author}
  {\bibfnamefont {W.}~\bibnamefont {van~der Schee}},\ and\ \bibinfo {author}
  {\bibfnamefont {U.~A.}\ \bibnamefont {Wiedemann}},\ }\bibfield  {title}
  {\bibinfo {title} {{Discovering Partonic Rescattering in Light Nucleus
  Collisions}},\ }\href {https://doi.org/10.1103/PhysRevLett.126.192301}
  {\bibfield  {journal} {\bibinfo  {journal} {Phys. Rev. Lett.}\ }\textbf
  {\bibinfo {volume} {126}},\ \bibinfo {pages} {192301} (\bibinfo {year}
  {2021}{\natexlab{a}})},\ \Eprint {https://arxiv.org/abs/2007.13754}
  {arXiv:2007.13754 [hep-ph]} \BibitemShut {NoStop}%
\bibitem [{\citenamefont {Huss}\ \emph
  {et~al.}(2021{\natexlab{b}})\citenamefont {Huss}, \citenamefont {Kurkela},
  \citenamefont {Mazeliauskas}, \citenamefont {Paatelainen}, \citenamefont
  {van~der Schee},\ and\ \citenamefont {Wiedemann}}]{Huss:2020whe}%
  \BibitemOpen
  \bibfield  {author} {\bibinfo {author} {\bibfnamefont {A.}~\bibnamefont
  {Huss}}, \bibinfo {author} {\bibfnamefont {A.}~\bibnamefont {Kurkela}},
  \bibinfo {author} {\bibfnamefont {A.}~\bibnamefont {Mazeliauskas}}, \bibinfo
  {author} {\bibfnamefont {R.}~\bibnamefont {Paatelainen}}, \bibinfo {author}
  {\bibfnamefont {W.}~\bibnamefont {van~der Schee}},\ and\ \bibinfo {author}
  {\bibfnamefont {U.~A.}\ \bibnamefont {Wiedemann}},\ }\bibfield  {title}
  {\bibinfo {title} {{Predicting parton energy loss in small collision
  systems}},\ }\href {https://doi.org/10.1103/PhysRevC.103.054903} {\bibfield
  {journal} {\bibinfo  {journal} {Phys. Rev. C}\ }\textbf {\bibinfo {volume}
  {103}},\ \bibinfo {pages} {054903} (\bibinfo {year} {2021}{\natexlab{b}})},\
  \Eprint {https://arxiv.org/abs/2007.13758} {arXiv:2007.13758 [hep-ph]}
  \BibitemShut {NoStop}%
\bibitem [{\citenamefont {Zakharov}(2021)}]{Zakharov:2021uza}%
  \BibitemOpen
  \bibfield  {author} {\bibinfo {author} {\bibfnamefont {B.~G.}\ \bibnamefont
  {Zakharov}},\ }\bibfield  {title} {\bibinfo {title} {{Jet quenching from
  heavy to light ion collisions}},\ }\href
  {https://doi.org/10.1007/JHEP09(2021)087} {\bibfield  {journal} {\bibinfo
  {journal} {JHEP}\ }\textbf {\bibinfo {volume} {09}},\ \bibinfo {pages}
  {087}},\ \Eprint {https://arxiv.org/abs/2105.09350} {arXiv:2105.09350
  [hep-ph]} \BibitemShut {NoStop}%
\bibitem [{\citenamefont {Ke}\ and\ \citenamefont {Vitev}(2023)}]{Ke:2022gkq}%
  \BibitemOpen
  \bibfield  {author} {\bibinfo {author} {\bibfnamefont {W.}~\bibnamefont
  {Ke}}\ and\ \bibinfo {author} {\bibfnamefont {I.}~\bibnamefont {Vitev}},\
  }\bibfield  {title} {\bibinfo {title} {{Searching for QGP droplets with
  high-pT hadrons and heavy flavor}},\ }\href
  {https://doi.org/10.1103/PhysRevC.107.064903} {\bibfield  {journal} {\bibinfo
   {journal} {Phys. Rev. C}\ }\textbf {\bibinfo {volume} {107}},\ \bibinfo
  {pages} {064903} (\bibinfo {year} {2023})},\ \Eprint
  {https://arxiv.org/abs/2204.00634} {arXiv:2204.00634 [hep-ph]} \BibitemShut
  {NoStop}%
\bibitem [{\citenamefont {Ogrodnik}\ \emph {et~al.}(2025)\citenamefont
  {Ogrodnik}, \citenamefont {Ryb{\'a}{\v{r}}},\ and\ \citenamefont
  {Spousta}}]{Ogrodnik:2024qug}%
  \BibitemOpen
  \bibfield  {author} {\bibinfo {author} {\bibfnamefont {A.}~\bibnamefont
  {Ogrodnik}}, \bibinfo {author} {\bibfnamefont {M.}~\bibnamefont
  {Ryb{\'a}{\v{r}}}},\ and\ \bibinfo {author} {\bibfnamefont {M.}~\bibnamefont
  {Spousta}},\ }\bibfield  {title} {\bibinfo {title} {{Flavor and path-length
  dependence of jet quenching from inclusive jet and $\gamma $-jet
  suppression}},\ }\href {https://doi.org/10.1140/epjc/s10052-025-14629-5}
  {\bibfield  {journal} {\bibinfo  {journal} {Eur. Phys. J. C}\ }\textbf
  {\bibinfo {volume} {85}},\ \bibinfo {pages} {899} (\bibinfo {year} {2025})},\
  \Eprint {https://arxiv.org/abs/2407.11234} {arXiv:2407.11234 [hep-ph]}
  \BibitemShut {NoStop}%
\bibitem [{\citenamefont {van~der Schee}\ \emph {et~al.}(2025)\citenamefont
  {van~der Schee}, \citenamefont {Kolb{\'e}}, \citenamefont {Nijs},
  \citenamefont {Ruhani}, \citenamefont {Ahmed},\ and\ \citenamefont
  {Iqbal}}]{vanderSchee:2025hoe}%
  \BibitemOpen
  \bibfield  {author} {\bibinfo {author} {\bibfnamefont {W.}~\bibnamefont
  {van~der Schee}}, \bibinfo {author} {\bibfnamefont {I.}~\bibnamefont
  {Kolb{\'e}}}, \bibinfo {author} {\bibfnamefont {G.}~\bibnamefont {Nijs}},
  \bibinfo {author} {\bibfnamefont {K.}~\bibnamefont {Ruhani}}, \bibinfo
  {author} {\bibfnamefont {I.}~\bibnamefont {Ahmed}},\ and\ \bibinfo {author}
  {\bibfnamefont {S.}~\bibnamefont {Iqbal}},\ }\bibfield  {title} {\bibinfo
  {title} {{Three models for charged hadron nuclear modification from light to
  heavy ions}},\ }\href@noop {} {\  (\bibinfo {year} {2025})},\ \Eprint
  {https://arxiv.org/abs/2509.04299} {arXiv:2509.04299 [nucl-th]} \BibitemShut
  {NoStop}%
\bibitem [{STA(2025)}]{STAR:2025ivi}%
  \BibitemOpen
  \bibfield  {title} {\bibinfo {title} {{Engineering the shapes of quark-gluon
  plasma droplets by comparing anisotropic flow in small symmetric and
  asymmetric collision systems}},\ }\href@noop {} {\  (\bibinfo {year}
  {2025})},\ \Eprint {https://arxiv.org/abs/2510.19645} {arXiv:2510.19645
  [nucl-ex]} \BibitemShut {NoStop}%
\bibitem [{\citenamefont {Belmont}\ \emph {et~al.}(2024)\citenamefont {Belmont}
  \emph {et~al.}}]{Belmont:2023fau}%
  \BibitemOpen
  \bibfield  {author} {\bibinfo {author} {\bibfnamefont {R.}~\bibnamefont
  {Belmont}} \emph {et~al.},\ }\bibfield  {title} {\bibinfo {title}
  {{Predictions for the sPHENIX physics program}},\ }\href
  {https://doi.org/10.1016/j.nuclphysa.2024.122821} {\bibfield  {journal}
  {\bibinfo  {journal} {Nucl. Phys. A}\ }\textbf {\bibinfo {volume} {1043}},\
  \bibinfo {pages} {122821} (\bibinfo {year} {2024})},\ \Eprint
  {https://arxiv.org/abs/2305.15491} {arXiv:2305.15491 [nucl-ex]} \BibitemShut
  {NoStop}%
\bibitem [{\citenamefont {Gebhard}\ \emph {et~al.}(2025)\citenamefont
  {Gebhard}, \citenamefont {Mazeliauskas},\ and\ \citenamefont
  {Takacs}}]{Gebhard:2024flv}%
  \BibitemOpen
  \bibfield  {author} {\bibinfo {author} {\bibfnamefont {J.}~\bibnamefont
  {Gebhard}}, \bibinfo {author} {\bibfnamefont {A.}~\bibnamefont
  {Mazeliauskas}},\ and\ \bibinfo {author} {\bibfnamefont {A.}~\bibnamefont
  {Takacs}},\ }\bibfield  {title} {\bibinfo {title} {{No-quenching baseline for
  energy loss signals in oxygen-oxygen collisions}},\ }\href
  {https://doi.org/10.1007/JHEP04(2025)034} {\bibfield  {journal} {\bibinfo
  {journal} {JHEP}\ }\textbf {\bibinfo {volume} {04}},\ \bibinfo {pages}
  {034}},\ \Eprint {https://arxiv.org/abs/2410.22405} {arXiv:2410.22405
  [hep-ph]} \BibitemShut {NoStop}%
\bibitem [{\citenamefont {Brewer}\ \emph {et~al.}(2022)\citenamefont {Brewer},
  \citenamefont {Huss}, \citenamefont {Mazeliauskas},\ and\ \citenamefont
  {van~der Schee}}]{Brewer:2021tyv}%
  \BibitemOpen
  \bibfield  {author} {\bibinfo {author} {\bibfnamefont {J.}~\bibnamefont
  {Brewer}}, \bibinfo {author} {\bibfnamefont {A.}~\bibnamefont {Huss}},
  \bibinfo {author} {\bibfnamefont {A.}~\bibnamefont {Mazeliauskas}},\ and\
  \bibinfo {author} {\bibfnamefont {W.}~\bibnamefont {van~der Schee}},\
  }\bibfield  {title} {\bibinfo {title} {{Ratios of jet and hadron spectra at
  LHC energies: Measuring high-$p_T$ suppression without a pp reference}},\
  }\href {https://doi.org/10.1103/PhysRevD.105.074040} {\bibfield  {journal}
  {\bibinfo  {journal} {Phys. Rev. D}\ }\textbf {\bibinfo {volume} {105}},\
  \bibinfo {pages} {074040} (\bibinfo {year} {2022})},\ \Eprint
  {https://arxiv.org/abs/2108.13434} {arXiv:2108.13434 [hep-ph]} \BibitemShut
  {NoStop}%
\bibitem [{\citenamefont {Adam}\ \emph {et~al.}(2015)\citenamefont {Adam} \emph
  {et~al.}}]{ALICE:2014xsp}%
  \BibitemOpen
  \bibfield  {author} {\bibinfo {author} {\bibfnamefont {J.}~\bibnamefont
  {Adam}} \emph {et~al.} (\bibinfo {collaboration} {ALICE}),\ }\bibfield
  {title} {\bibinfo {title} {{Centrality dependence of particle production in
  p-Pb collisions at $\sqrt{s_{\rm NN} }$= 5.02 TeV}},\ }\href
  {https://doi.org/10.1103/PhysRevC.91.064905} {\bibfield  {journal} {\bibinfo
  {journal} {Phys. Rev. C}\ }\textbf {\bibinfo {volume} {91}},\ \bibinfo
  {pages} {064905} (\bibinfo {year} {2015})},\ \Eprint
  {https://arxiv.org/abs/1412.6828} {arXiv:1412.6828 [nucl-ex]} \BibitemShut
  {NoStop}%
\bibitem [{\citenamefont {Klein}\ and\ \citenamefont
  {Steinberg}(2020)}]{Klein:2020fmr}%
  \BibitemOpen
  \bibfield  {author} {\bibinfo {author} {\bibfnamefont {S.}~\bibnamefont
  {Klein}}\ and\ \bibinfo {author} {\bibfnamefont {P.}~\bibnamefont
  {Steinberg}},\ }\bibfield  {title} {\bibinfo {title} {{Photonuclear and
  Two-photon Interactions at High-Energy Nuclear Colliders}},\ }\href
  {https://doi.org/10.1146/annurev-nucl-030320-033923} {\bibfield  {journal}
  {\bibinfo  {journal} {Ann. Rev. Nucl. Part. Sci.}\ }\textbf {\bibinfo
  {volume} {70}},\ \bibinfo {pages} {323} (\bibinfo {year} {2020})},\ \Eprint
  {https://arxiv.org/abs/2005.01872} {arXiv:2005.01872 [nucl-ex]} \BibitemShut
  {NoStop}%
\bibitem [{\citenamefont {Bruce}\ \emph {et~al.}(2021)\citenamefont {Bruce},
  \citenamefont {Alemany-Fern{\'a}ndez}, \citenamefont {Bartosik},
  \citenamefont {Jebramcik}, \citenamefont {Jowett},\ and\ \citenamefont
  {Schaumann}}]{Bruce:2021hjk}%
  \BibitemOpen
  \bibfield  {author} {\bibinfo {author} {\bibfnamefont {R.}~\bibnamefont
  {Bruce}}, \bibinfo {author} {\bibfnamefont {R.}~\bibnamefont
  {Alemany-Fern{\'a}ndez}}, \bibinfo {author} {\bibfnamefont {H.}~\bibnamefont
  {Bartosik}}, \bibinfo {author} {\bibfnamefont {M.}~\bibnamefont {Jebramcik}},
  \bibinfo {author} {\bibfnamefont {J.}~\bibnamefont {Jowett}},\ and\ \bibinfo
  {author} {\bibfnamefont {M.}~\bibnamefont {Schaumann}},\ }\bibfield  {title}
  {\bibinfo {title} {{Studies for an LHC Pilot Run with Oxygen Beams}},\ }in\
  \href {https://doi.org/10.18429/JACoW-IPAC2021-MOPAB005} {\emph {\bibinfo
  {booktitle} {{12th International Particle Accelerator Conference~}}}}\
  (\bibinfo {year} {2021})\BibitemShut {NoStop}%
\bibitem [{\citenamefont {Nijs}\ and\ \citenamefont {van~der
  Schee}(2025)}]{Nijs:2025qxm}%
  \BibitemOpen
  \bibfield  {author} {\bibinfo {author} {\bibfnamefont {G.}~\bibnamefont
  {Nijs}}\ and\ \bibinfo {author} {\bibfnamefont {W.}~\bibnamefont {van~der
  Schee}},\ }\bibfield  {title} {\bibinfo {title} {{Transmutation of $^{16}$O
  and $^{20}$Ne at the Large Hadron Collider}},\ }\href@noop {} {\  (\bibinfo
  {year} {2025})},\ \Eprint {https://arxiv.org/abs/2507.01659}
  {arXiv:2507.01659 [nucl-th]} \BibitemShut {NoStop}%
\bibitem [{\citenamefont {Acharya}\ \emph
  {et~al.}(2019{\natexlab{a}})\citenamefont {Acharya} \emph
  {et~al.}}]{ALICE:2019fhe}%
  \BibitemOpen
  \bibfield  {author} {\bibinfo {author} {\bibfnamefont {S.}~\bibnamefont
  {Acharya}} \emph {et~al.} (\bibinfo {collaboration} {ALICE}),\ }\bibfield
  {title} {\bibinfo {title} {{Measurement of prompt D$^{0}$, D$^{+}$, D$^{*+}$,
  and $ {\mathrm{D}}_{\mathrm{S}}^{+} $ production in p{\textendash}Pb
  collisions at $ \sqrt{{\mathrm{s}}_{\mathrm{NN}}} $ = 5.02 TeV}},\ }\href
  {https://doi.org/10.1007/JHEP12(2019)092} {\bibfield  {journal} {\bibinfo
  {journal} {JHEP}\ }\textbf {\bibinfo {volume} {12}},\ \bibinfo {pages}
  {092}},\ \Eprint {https://arxiv.org/abs/1906.03425} {arXiv:1906.03425
  [nucl-ex]} \BibitemShut {NoStop}%
\bibitem [{\citenamefont {Adare}\ \emph {et~al.}(2014)\citenamefont {Adare}
  \emph {et~al.}}]{PHENIX:2013jxf}%
  \BibitemOpen
  \bibfield  {author} {\bibinfo {author} {\bibfnamefont {A.}~\bibnamefont
  {Adare}} \emph {et~al.} (\bibinfo {collaboration} {PHENIX}),\ }\bibfield
  {title} {\bibinfo {title} {{Centrality categorization for $R_{p(d)+A}$ in
  high-energy collisions}},\ }\href
  {https://doi.org/10.1103/PhysRevC.90.034902} {\bibfield  {journal} {\bibinfo
  {journal} {Phys. Rev. C}\ }\textbf {\bibinfo {volume} {90}},\ \bibinfo
  {pages} {034902} (\bibinfo {year} {2014})},\ \Eprint
  {https://arxiv.org/abs/1310.4793} {arXiv:1310.4793 [nucl-ex]} \BibitemShut
  {NoStop}%
\bibitem [{\citenamefont {Perepelitsa}\ and\ \citenamefont
  {Steinberg}(2014)}]{Perepelitsa:2014yta}%
  \BibitemOpen
  \bibfield  {author} {\bibinfo {author} {\bibfnamefont {D.~V.}\ \bibnamefont
  {Perepelitsa}}\ and\ \bibinfo {author} {\bibfnamefont {P.~A.}\ \bibnamefont
  {Steinberg}},\ }\bibfield  {title} {\bibinfo {title} {{Calculation of
  centrality bias factors in $p$+A collisions based on a positive correlation
  of hard process yields with underlying event activity}},\ }\href@noop {} {\
  (\bibinfo {year} {2014})},\ \Eprint {https://arxiv.org/abs/1412.0976}
  {arXiv:1412.0976 [nucl-ex]} \BibitemShut {NoStop}%
\bibitem [{\citenamefont {Loizides}\ and\ \citenamefont
  {Morsch}(2017)}]{Loizides:2017sqq}%
  \BibitemOpen
  \bibfield  {author} {\bibinfo {author} {\bibfnamefont {C.}~\bibnamefont
  {Loizides}}\ and\ \bibinfo {author} {\bibfnamefont {A.}~\bibnamefont
  {Morsch}},\ }\bibfield  {title} {\bibinfo {title} {{Absence of jet quenching
  in peripheral nucleus\textendash{}nucleus collisions}},\ }\href
  {https://doi.org/10.1016/j.physletb.2017.09.002} {\bibfield  {journal}
  {\bibinfo  {journal} {Phys. Lett. B}\ }\textbf {\bibinfo {volume} {773}},\
  \bibinfo {pages} {408} (\bibinfo {year} {2017})},\ \Eprint
  {https://arxiv.org/abs/1705.08856} {arXiv:1705.08856 [nucl-ex]} \BibitemShut
  {NoStop}%
\bibitem [{\citenamefont {Sjostrand}\ and\ \citenamefont {van
  Zijl}(1987)}]{Sjostrand:1986ep}%
  \BibitemOpen
  \bibfield  {author} {\bibinfo {author} {\bibfnamefont {T.}~\bibnamefont
  {Sjostrand}}\ and\ \bibinfo {author} {\bibfnamefont {M.}~\bibnamefont {van
  Zijl}},\ }\bibfield  {title} {\bibinfo {title} {{Multiple Parton-parton
  Interactions in an Impact Parameter Picture}},\ }\href
  {https://doi.org/10.1016/0370-2693(87)90722-2} {\bibfield  {journal}
  {\bibinfo  {journal} {Phys. Lett. B}\ }\textbf {\bibinfo {volume} {188}},\
  \bibinfo {pages} {149} (\bibinfo {year} {1987})}\BibitemShut {NoStop}%
\bibitem [{\citenamefont {Jia}(2009)}]{Jia:2009mq}%
  \BibitemOpen
  \bibfield  {author} {\bibinfo {author} {\bibfnamefont {J.}~\bibnamefont
  {Jia}},\ }\bibfield  {title} {\bibinfo {title} {{Influence of the
  nucleon-nucleon collision geometry on the determination of the nuclear
  modification factor for nucleon-nucleus and nucleus-nucleus collisions}},\
  }\href {https://doi.org/10.1016/j.physletb.2009.10.044} {\bibfield  {journal}
  {\bibinfo  {journal} {Phys. Lett. B}\ }\textbf {\bibinfo {volume} {681}},\
  \bibinfo {pages} {320} (\bibinfo {year} {2009})},\ \Eprint
  {https://arxiv.org/abs/0907.4175} {arXiv:0907.4175 [nucl-th]} \BibitemShut
  {NoStop}%
\bibitem [{\citenamefont {Frankfurt}\ \emph {et~al.}(2011)\citenamefont
  {Frankfurt}, \citenamefont {Strikman},\ and\ \citenamefont
  {Weiss}}]{Frankfurt:2010ea}%
  \BibitemOpen
  \bibfield  {author} {\bibinfo {author} {\bibfnamefont {L.}~\bibnamefont
  {Frankfurt}}, \bibinfo {author} {\bibfnamefont {M.}~\bibnamefont
  {Strikman}},\ and\ \bibinfo {author} {\bibfnamefont {C.}~\bibnamefont
  {Weiss}},\ }\bibfield  {title} {\bibinfo {title} {{Transverse nucleon
  structure and diagnostics of hard parton-parton processes at LHC}},\ }\href
  {https://doi.org/10.1103/PhysRevD.83.054012} {\bibfield  {journal} {\bibinfo
  {journal} {Phys. Rev. D}\ }\textbf {\bibinfo {volume} {83}},\ \bibinfo
  {pages} {054012} (\bibinfo {year} {2011})},\ \Eprint
  {https://arxiv.org/abs/1009.2559} {arXiv:1009.2559 [hep-ph]} \BibitemShut
  {NoStop}%
\bibitem [{\citenamefont {Affolder}\ \emph {et~al.}(2002)\citenamefont
  {Affolder} \emph {et~al.}}]{CDF:2001onq}%
  \BibitemOpen
  \bibfield  {author} {\bibinfo {author} {\bibfnamefont {T.}~\bibnamefont
  {Affolder}} \emph {et~al.} (\bibinfo {collaboration} {CDF}),\ }\bibfield
  {title} {\bibinfo {title} {{Charged Jet Evolution and the Underlying Event in
  $p\bar{p}$ Collisions at 1.8 TeV}},\ }\href
  {https://doi.org/10.1103/PhysRevD.65.092002} {\bibfield  {journal} {\bibinfo
  {journal} {Phys. Rev. D}\ }\textbf {\bibinfo {volume} {65}},\ \bibinfo
  {pages} {092002} (\bibinfo {year} {2002})}\BibitemShut {NoStop}%
\bibitem [{\citenamefont {Acharya}\ \emph
  {et~al.}(2019{\natexlab{b}})\citenamefont {Acharya} \emph
  {et~al.}}]{ALICE:2018ekf}%
  \BibitemOpen
  \bibfield  {author} {\bibinfo {author} {\bibfnamefont {S.}~\bibnamefont
  {Acharya}} \emph {et~al.} (\bibinfo {collaboration} {ALICE}),\ }\bibfield
  {title} {\bibinfo {title} {{Analysis of the apparent nuclear modification in
  peripheral Pb\textendash{}Pb collisions at 5.02 TeV}},\ }\href
  {https://doi.org/10.1016/j.physletb.2019.04.047} {\bibfield  {journal}
  {\bibinfo  {journal} {Phys. Lett. B}\ }\textbf {\bibinfo {volume} {793}},\
  \bibinfo {pages} {420} (\bibinfo {year} {2019}{\natexlab{b}})},\ \Eprint
  {https://arxiv.org/abs/1805.05212} {arXiv:1805.05212 [nucl-ex]} \BibitemShut
  {NoStop}%
\bibitem [{\citenamefont {Sirunyan}\ \emph {et~al.}(2021)\citenamefont
  {Sirunyan} \emph {et~al.}}]{CMS:2021kvd}%
  \BibitemOpen
  \bibfield  {author} {\bibinfo {author} {\bibfnamefont {A.~M.}\ \bibnamefont
  {Sirunyan}} \emph {et~al.} (\bibinfo {collaboration} {CMS}),\ }\bibfield
  {title} {\bibinfo {title} {{Constraints on the Initial State of Pb-Pb
  Collisions via Measurements of $Z$-Boson Yields and Azimuthal Anisotropy at
  $\sqrt {s_{NN}}$=5.02\,\,TeV}},\ }\href
  {https://doi.org/10.1103/PhysRevLett.127.102002} {\bibfield  {journal}
  {\bibinfo  {journal} {Phys. Rev. Lett.}\ }\textbf {\bibinfo {volume} {127}},\
  \bibinfo {pages} {102002} (\bibinfo {year} {2021})},\ \Eprint
  {https://arxiv.org/abs/2103.14089} {arXiv:2103.14089 [hep-ex]} \BibitemShut
  {NoStop}%
\bibitem [{\citenamefont {Acharya}\ \emph
  {et~al.}(2025{\natexlab{a}})\citenamefont {Acharya} \emph
  {et~al.}}]{ALICE:2024yvg}%
  \BibitemOpen
  \bibfield  {author} {\bibinfo {author} {\bibfnamefont {S.}~\bibnamefont
  {Acharya}} \emph {et~al.} (\bibinfo {collaboration} {ALICE}),\ }\bibfield
  {title} {\bibinfo {title} {{Measurement of the inclusive isolated-photon
  production cross section in pp and Pb{\textendash}Pb collisions at $\mathbf
  {\sqrt{\textit{s}_{NN }} = 5.02}$~TeV}},\ }\href
  {https://doi.org/10.1140/epjc/s10052-025-13971-y} {\bibfield  {journal}
  {\bibinfo  {journal} {Eur. Phys. J. C}\ }\textbf {\bibinfo {volume} {85}},\
  \bibinfo {pages} {553} (\bibinfo {year} {2025}{\natexlab{a}})},\ \Eprint
  {https://arxiv.org/abs/2409.12641} {arXiv:2409.12641 [nucl-ex]} \BibitemShut
  {NoStop}%
\bibitem [{\citenamefont {Sjostrand}\ \emph {et~al.}(2006)\citenamefont
  {Sjostrand}, \citenamefont {Mrenna},\ and\ \citenamefont
  {Skands}}]{Sjostrand:2006za}%
  \BibitemOpen
  \bibfield  {author} {\bibinfo {author} {\bibfnamefont {T.}~\bibnamefont
  {Sjostrand}}, \bibinfo {author} {\bibfnamefont {S.}~\bibnamefont {Mrenna}},\
  and\ \bibinfo {author} {\bibfnamefont {P.~Z.}\ \bibnamefont {Skands}},\
  }\bibfield  {title} {\bibinfo {title} {{PYTHIA 6.4 Physics and Manual}},\
  }\href {https://doi.org/10.1088/1126-6708/2006/05/026} {\bibfield  {journal}
  {\bibinfo  {journal} {JHEP}\ }\textbf {\bibinfo {volume} {05}},\ \bibinfo
  {pages} {026}},\ \Eprint {https://arxiv.org/abs/hep-ph/0603175}
  {arXiv:hep-ph/0603175} \BibitemShut {NoStop}%
\bibitem [{\citenamefont {Wang}\ and\ \citenamefont
  {Gyulassy}(1991)}]{Wang:1991hta}%
  \BibitemOpen
  \bibfield  {author} {\bibinfo {author} {\bibfnamefont {X.-N.}\ \bibnamefont
  {Wang}}\ and\ \bibinfo {author} {\bibfnamefont {M.}~\bibnamefont
  {Gyulassy}},\ }\bibfield  {title} {\bibinfo {title} {{HIJING: A Monte Carlo
  model for multiple jet production in p p, p A and A A collisions}},\ }\href
  {https://doi.org/10.1103/PhysRevD.44.3501} {\bibfield  {journal} {\bibinfo
  {journal} {Phys. Rev. D}\ }\textbf {\bibinfo {volume} {44}},\ \bibinfo
  {pages} {3501} (\bibinfo {year} {1991})}\BibitemShut {NoStop}%
\bibitem [{\citenamefont {Bierlich}\ \emph {et~al.}(2018)\citenamefont
  {Bierlich}, \citenamefont {Gustafson}, \citenamefont {L\"onnblad},\ and\
  \citenamefont {Shah}}]{Bierlich:2018xfw}%
  \BibitemOpen
  \bibfield  {author} {\bibinfo {author} {\bibfnamefont {C.}~\bibnamefont
  {Bierlich}}, \bibinfo {author} {\bibfnamefont {G.}~\bibnamefont {Gustafson}},
  \bibinfo {author} {\bibfnamefont {L.}~\bibnamefont {L\"onnblad}},\ and\
  \bibinfo {author} {\bibfnamefont {H.}~\bibnamefont {Shah}},\ }\bibfield
  {title} {\bibinfo {title} {{The Angantyr model for Heavy-Ion Collisions in
  PYTHIA8}},\ }\href {https://doi.org/10.1007/JHEP10(2018)134} {\bibfield
  {journal} {\bibinfo  {journal} {JHEP}\ }\textbf {\bibinfo {volume} {10}},\
  \bibinfo {pages} {134}},\ \Eprint {https://arxiv.org/abs/1806.10820}
  {arXiv:1806.10820 [hep-ph]} \BibitemShut {NoStop}%
\bibitem [{\citenamefont {Bierlich}\ \emph {et~al.}(2022)\citenamefont
  {Bierlich} \emph {et~al.}}]{Bierlich:2022pfr}%
  \BibitemOpen
  \bibfield  {author} {\bibinfo {author} {\bibfnamefont {C.}~\bibnamefont
  {Bierlich}} \emph {et~al.},\ }\bibfield  {title} {\bibinfo {title} {{A
  comprehensive guide to the physics and usage of PYTHIA 8.3}},\ }\href
  {https://doi.org/10.21468/SciPostPhysCodeb.8} {\bibfield  {journal} {\bibinfo
   {journal} {SciPost Phys. Codeb.}\ }\textbf {\bibinfo {volume} {2022}},\
  \bibinfo {pages} {8} (\bibinfo {year} {2022})},\ \Eprint
  {https://arxiv.org/abs/2203.11601} {arXiv:2203.11601 [hep-ph]} \BibitemShut
  {NoStop}%
\bibitem [{\citenamefont {Lim}\ \emph {et~al.}(2019)\citenamefont {Lim},
  \citenamefont {Carlson}, \citenamefont {Loizides}, \citenamefont {Lonardoni},
  \citenamefont {Lynn}, \citenamefont {Nagle}, \citenamefont {Orjuela~Koop},\
  and\ \citenamefont {Ouellette}}]{Lim:2018huo}%
  \BibitemOpen
  \bibfield  {author} {\bibinfo {author} {\bibfnamefont {S.~H.}\ \bibnamefont
  {Lim}}, \bibinfo {author} {\bibfnamefont {J.}~\bibnamefont {Carlson}},
  \bibinfo {author} {\bibfnamefont {C.}~\bibnamefont {Loizides}}, \bibinfo
  {author} {\bibfnamefont {D.}~\bibnamefont {Lonardoni}}, \bibinfo {author}
  {\bibfnamefont {J.~E.}\ \bibnamefont {Lynn}}, \bibinfo {author}
  {\bibfnamefont {J.~L.}\ \bibnamefont {Nagle}}, \bibinfo {author}
  {\bibfnamefont {J.~D.}\ \bibnamefont {Orjuela~Koop}},\ and\ \bibinfo {author}
  {\bibfnamefont {J.}~\bibnamefont {Ouellette}},\ }\bibfield  {title} {\bibinfo
  {title} {{Exploring New Small System Geometries in Heavy Ion Collisions}},\
  }\href {https://doi.org/10.1103/PhysRevC.99.044904} {\bibfield  {journal}
  {\bibinfo  {journal} {Phys. Rev. C}\ }\textbf {\bibinfo {volume} {99}},\
  \bibinfo {pages} {044904} (\bibinfo {year} {2019})},\ \Eprint
  {https://arxiv.org/abs/1812.08096} {arXiv:1812.08096 [nucl-th]} \BibitemShut
  {NoStop}%
\bibitem [{\citenamefont {Adams}\ \emph {et~al.}(2020)\citenamefont {Adams}
  \emph {et~al.}}]{Adams:2019fpo}%
  \BibitemOpen
  \bibfield  {author} {\bibinfo {author} {\bibfnamefont {J.}~\bibnamefont
  {Adams}} \emph {et~al.},\ }\bibfield  {title} {\bibinfo {title} {{The STAR
  Event Plane Detector}},\ }\href {https://doi.org/10.1016/j.nima.2020.163970}
  {\bibfield  {journal} {\bibinfo  {journal} {Nucl. Instrum. Meth. A}\ }\textbf
  {\bibinfo {volume} {968}},\ \bibinfo {pages} {163970} (\bibinfo {year}
  {2020})},\ \Eprint {https://arxiv.org/abs/1912.05243} {arXiv:1912.05243
  [physics.ins-det]} \BibitemShut {NoStop}%
\bibitem [{\citenamefont {Aad}\ \emph {et~al.}(2016{\natexlab{a}})\citenamefont
  {Aad} \emph {et~al.}}]{ATLAS:2015hkr}%
  \BibitemOpen
  \bibfield  {author} {\bibinfo {author} {\bibfnamefont {G.}~\bibnamefont
  {Aad}} \emph {et~al.} (\bibinfo {collaboration} {ATLAS}),\ }\bibfield
  {title} {\bibinfo {title} {{Measurement of the centrality dependence of the
  charged-particle pseudorapidity distribution in proton\textendash{}lead
  collisions at $\sqrt{s_{_\text {NN}}} = 5.02$ TeV with the ATLAS detector}},\
  }\href {https://doi.org/10.1140/epjc/s10052-016-4002-3} {\bibfield  {journal}
  {\bibinfo  {journal} {Eur. Phys. J. C}\ }\textbf {\bibinfo {volume} {76}},\
  \bibinfo {pages} {199} (\bibinfo {year} {2016}{\natexlab{a}})},\ \Eprint
  {https://arxiv.org/abs/1508.00848} {arXiv:1508.00848 [hep-ex]} \BibitemShut
  {NoStop}%
\bibitem [{\citenamefont {Sirunyan}\ \emph {et~al.}(2019)\citenamefont
  {Sirunyan} \emph {et~al.}}]{CMS:2018xfv}%
  \BibitemOpen
  \bibfield  {author} {\bibinfo {author} {\bibfnamefont {A.~M.}\ \bibnamefont
  {Sirunyan}} \emph {et~al.} (\bibinfo {collaboration} {CMS}),\ }\bibfield
  {title} {\bibinfo {title} {{Centrality and pseudorapidity dependence of the
  transverse energy density in pPb collisions at $\sqrt{s_\mathrm{NN}} =$ 5.02
  TeV}},\ }\href {https://doi.org/10.1103/PhysRevC.100.024902} {\bibfield
  {journal} {\bibinfo  {journal} {Phys. Rev. C}\ }\textbf {\bibinfo {volume}
  {100}},\ \bibinfo {pages} {024902} (\bibinfo {year} {2019})},\ \Eprint
  {https://arxiv.org/abs/1810.05745} {arXiv:1810.05745 [hep-ex]} \BibitemShut
  {NoStop}%
\bibitem [{\citenamefont {Acharya}\ \emph
  {et~al.}(2025{\natexlab{b}})\citenamefont {Acharya} \emph
  {et~al.}}]{ALICE:2025cjn}%
  \BibitemOpen
  \bibfield  {author} {\bibinfo {author} {\bibfnamefont {S.}~\bibnamefont
  {Acharya}} \emph {et~al.} (\bibinfo {collaboration} {ALICE}),\ }\bibfield
  {title} {\bibinfo {title} {{Centrality dependence of charged-particle
  pseudorapidity density at midrapidity in Pb-Pb collisions at
  $\mathbf{\sqrt{\textit{s}_{\rm NN}} = 5.36}$ TeV}},\ }\href@noop {} {\
  (\bibinfo {year} {2025}{\natexlab{b}})},\ \Eprint
  {https://arxiv.org/abs/2504.02505} {arXiv:2504.02505 [nucl-ex]} \BibitemShut
  {NoStop}%
\bibitem [{\citenamefont {Abelev}\ \emph {et~al.}(2013)\citenamefont {Abelev}
  \emph {et~al.}}]{ALICE:2013hur}%
  \BibitemOpen
  \bibfield  {author} {\bibinfo {author} {\bibfnamefont {B.}~\bibnamefont
  {Abelev}} \emph {et~al.} (\bibinfo {collaboration} {ALICE}),\ }\bibfield
  {title} {\bibinfo {title} {{Centrality determination of Pb-Pb collisions at
  $\sqrt{s_{NN}}$ = 2.76 TeV with ALICE}},\ }\href
  {https://doi.org/10.1103/PhysRevC.88.044909} {\bibfield  {journal} {\bibinfo
  {journal} {Phys. Rev. C}\ }\textbf {\bibinfo {volume} {88}},\ \bibinfo
  {pages} {044909} (\bibinfo {year} {2013})},\ \Eprint
  {https://arxiv.org/abs/1301.4361} {arXiv:1301.4361 [nucl-ex]} \BibitemShut
  {NoStop}%
\bibitem [{\citenamefont {Zhang}(2025)}]{Zhang:2025cgr}%
  \BibitemOpen
  \bibfield  {author} {\bibinfo {author} {\bibfnamefont {S.}~\bibnamefont
  {Zhang}},\ }\bibfield  {title} {\bibinfo {title} {{Studies of jet quenching
  in O+O collisions at $\sqrt{s_{\rm NN}}$ = 200 GeV by STAR}},\ }in\
  \href@noop {} {\emph {\bibinfo {booktitle} {{31st International Conference on
  Ultra-relativistic Nucleus-Nucleus Collisions}}}}\ (\bibinfo {year} {2025})\
  \Eprint {https://arxiv.org/abs/2510.18616} {arXiv:2510.18616 [nucl-ex]}
  \BibitemShut {NoStop}%
\bibitem [{\citenamefont {Aad}\ \emph {et~al.}(2023{\natexlab{b}})\citenamefont
  {Aad} \emph {et~al.}}]{ATLAS:2022kqu}%
  \BibitemOpen
  \bibfield  {author} {\bibinfo {author} {\bibfnamefont {G.}~\bibnamefont
  {Aad}} \emph {et~al.} (\bibinfo {collaboration} {ATLAS}),\ }\bibfield
  {title} {\bibinfo {title} {{Charged-hadron production in $pp$, $p$+Pb, Pb+Pb,
  and Xe+Xe collisions at $\sqrt{s_{_\text{NN}}}=5$ TeV with the ATLAS detector
  at the LHC}},\ }\href {https://doi.org/10.1007/JHEP07(2023)074} {\bibfield
  {journal} {\bibinfo  {journal} {JHEP}\ }\textbf {\bibinfo {volume} {07}},\
  \bibinfo {pages} {074}},\ \Eprint {https://arxiv.org/abs/2211.15257}
  {arXiv:2211.15257 [hep-ex]} \BibitemShut {NoStop}%
\bibitem [{\citenamefont {Aad}\ \emph {et~al.}(2023{\natexlab{c}})\citenamefont
  {Aad} \emph {et~al.}}]{ATLAS:2022vii}%
  \BibitemOpen
  \bibfield  {author} {\bibinfo {author} {\bibfnamefont {G.}~\bibnamefont
  {Aad}} \emph {et~al.} (\bibinfo {collaboration} {ATLAS}),\ }\bibfield
  {title} {\bibinfo {title} {{Measurement of substructure-dependent jet
  suppression in Pb+Pb collisions at 5.02 TeV with the ATLAS detector}},\
  }\href {https://doi.org/10.1103/PhysRevC.107.054909} {\bibfield  {journal}
  {\bibinfo  {journal} {Phys. Rev. C}\ }\textbf {\bibinfo {volume} {107}},\
  \bibinfo {pages} {054909} (\bibinfo {year} {2023}{\natexlab{c}})},\ \Eprint
  {https://arxiv.org/abs/2211.11470} {arXiv:2211.11470 [nucl-ex]} \BibitemShut
  {NoStop}%
\bibitem [{\citenamefont {Aad}\ \emph {et~al.}(2016{\natexlab{b}})\citenamefont
  {Aad} \emph {et~al.}}]{ATLAS:2015vql}%
  \BibitemOpen
  \bibfield  {author} {\bibinfo {author} {\bibfnamefont {G.}~\bibnamefont
  {Aad}} \emph {et~al.} (\bibinfo {collaboration} {ATLAS}),\ }\bibfield
  {title} {\bibinfo {title} {{Measurement of the dependence of transverse
  energy production at large pseudorapidity on the hard-scattering kinematics
  of proton-proton collisions at $\sqrt{s} = 2.76$ TeV with ATLAS}},\ }\href
  {https://doi.org/10.1016/j.physletb.2016.02.056} {\bibfield  {journal}
  {\bibinfo  {journal} {Phys. Lett. B}\ }\textbf {\bibinfo {volume} {756}},\
  \bibinfo {pages} {10} (\bibinfo {year} {2016}{\natexlab{b}})},\ \Eprint
  {https://arxiv.org/abs/1512.00197} {arXiv:1512.00197 [hep-ex]} \BibitemShut
  {NoStop}%
\bibitem [{\citenamefont {Jonas}\ and\ \citenamefont
  {Loizides}(2021)}]{Jonas:2021xju}%
  \BibitemOpen
  \bibfield  {author} {\bibinfo {author} {\bibfnamefont {F.}~\bibnamefont
  {Jonas}}\ and\ \bibinfo {author} {\bibfnamefont {C.}~\bibnamefont
  {Loizides}},\ }\bibfield  {title} {\bibinfo {title} {{Centrality dependence
  of electroweak boson production in PbPb collisions at the CERN Large Hadron
  Collider}},\ }\href {https://doi.org/10.1103/PhysRevC.104.044905} {\bibfield
  {journal} {\bibinfo  {journal} {Phys. Rev. C}\ }\textbf {\bibinfo {volume}
  {104}},\ \bibinfo {pages} {044905} (\bibinfo {year} {2021})},\ \Eprint
  {https://arxiv.org/abs/2104.14903} {arXiv:2104.14903 [nucl-ex]} \BibitemShut
  {NoStop}%
\bibitem [{\citenamefont {Agostinelli}\ \emph {et~al.}(2003)\citenamefont
  {Agostinelli} \emph {et~al.}}]{GEANT4:2002zbu}%
  \BibitemOpen
  \bibfield  {author} {\bibinfo {author} {\bibfnamefont {S.}~\bibnamefont
  {Agostinelli}} \emph {et~al.} (\bibinfo {collaboration} {GEANT4}),\
  }\bibfield  {title} {\bibinfo {title} {{GEANT4 - A Simulation Toolkit}},\
  }\href {https://doi.org/10.1016/S0168-9002(03)01368-8} {\bibfield  {journal}
  {\bibinfo  {journal} {Nucl. Instrum. Meth. A}\ }\textbf {\bibinfo {volume}
  {506}},\ \bibinfo {pages} {250} (\bibinfo {year} {2003})}\BibitemShut
  {NoStop}%
\bibitem [{\citenamefont {Acharya}\ \emph {et~al.}(2024)\citenamefont {Acharya}
  \emph {et~al.}}]{ALICE:2023plt}%
  \BibitemOpen
  \bibfield  {author} {\bibinfo {author} {\bibfnamefont {S.}~\bibnamefont
  {Acharya}} \emph {et~al.} (\bibinfo {collaboration} {ALICE}),\ }\bibfield
  {title} {\bibinfo {title} {{Search for jet quenching effects in
  high-multiplicity pp collisions at $ \sqrt{s} $ = 13 TeV via di-jet
  acoplanarity}},\ }\href {https://doi.org/10.1007/JHEP05(2024)229} {\bibfield
  {journal} {\bibinfo  {journal} {JHEP}\ }\textbf {\bibinfo {volume} {05}},\
  \bibinfo {pages} {229}},\ \Eprint {https://arxiv.org/abs/2309.03788}
  {arXiv:2309.03788 [hep-ex]} \BibitemShut {NoStop}%
\end{thebibliography}%

\end{document}